\documentclass[journal]{IEEEtran}

\usepackage{scalerel}
\usepackage{tikz}
\usetikzlibrary{svg.path}
\usepackage{xcolor}
\usepackage{amsmath}
\usepackage{amsfonts}
\usepackage{mathtools}
\usepackage[numbers]{natbib}
\usepackage[ruled,vlined,noend,linesnumbered]{algorithm2e}
\usepackage{caption}
\usepackage{subcaption}
\usepackage{multirow}
\usepackage{booktabs}
\usepackage{blkarray}
\usepackage[utf8]{inputenc}
\usepackage{fancyhdr}
\usepackage{pifont}
\newcommand{\cmark}{\ding{51}}%

\definecolor{mac}{HTML}{0000FF}
\definecolor{sftm}{HTML}{008800}
\definecolor{ln}{HTML}{FF8C00}

\newcommand{\code}[1]{\textbf{\texttt{#1}}}

\definecolor{orcidlogocol}{HTML}{A6CE39}
\tikzset{
    orcidlogo/.pic={
        \fill[orcidlogocol] svg{M256,128c0,70.7-57.3,128-128,128C57.3,256,0,198.7,0,128C0,57.3,57.3,0,128,0C198.7,0,256,57.3,256,128z};
        \fill[white] svg{M86.3,186.2H70.9V79.1h15.4v48.4V186.2z}
        svg{M108.9,79.1h41.6c39.6,0,57,28.3,57,53.6c0,27.5-21.5,53.6-56.8,53.6h-41.8V79.1z M124.3,172.4h24.5c34.9,0,42.9-26.5,42.9-39.7c0-21.5-13.7-39.7-43.7-39.7h-23.7V172.4z}
        svg{M88.7,56.8c0,5.5-4.5,10.1-10.1,10.1c-5.6,0-10.1-4.6-10.1-10.1c0-5.6,4.5-10.1,10.1-10.1C84.2,46.7,88.7,51.3,88.7,56.8z};
    }
}

\newcommand\orcidicon[1]{\href{https://orcid.org/#1}{\mbox{\scalerel*{
                \begin{tikzpicture}[yscale=-1,transform shape]
                \pic{orcidlogo};
                \end{tikzpicture}
            }{|}}}}

\usepackage{hyperref}

\def\@fnsymbol#1{\ensuremath{\ifcase#1\or *\or \dagger\or \ddagger\or
   \mathsection\or \mathparagraph\or \|\or **\or \dagger\dagger
   \or \ddagger\ddagger \else\@ctrerr\fi}}
\newcommand{\ssymbol}[1]{^{\@fnsymbol{#1}}}

\begin{document}
\title{AccelTran: A Sparsity-Aware Accelerator for Dynamic Inference with Transformers}

\author{Shikhar~Tuli$^{\textsuperscript{\orcidicon{0000-0002-9230-5877}}}$,~\IEEEmembership{Student Member,~IEEE,} and~Niraj~K.~Jha,~\IEEEmembership{Fellow,~IEEE}
\thanks{This work was supported by NSF Grant No. CCF-2203399. S. Tuli and N. K. Jha are with the Department 
of Electrical and Computer Engineering,
Princeton University, Princeton, NJ, 08544, USA (e-mail: \{stuli, jha\}@princeton.edu).}
\thanks{Manuscript received ---; revised ---.}}

\markboth{}{Tuli \MakeLowercase{\textit{et al.}}: AccelTran: A Sparsity-Aware Accelerator for Dynamic Inference of Transformers}


\maketitle

\begin{abstract}

Self-attention-based transformer models have achieved tremendous success in the domain of natural language processing.
Despite their efficacy, accelerating the transformer is challenging due to its quadratic computational
complexity and large activation sizes. Existing transformer accelerators attempt to prune its tokens to reduce memory access, albeit with high compute overheads. Moreover, previous works directly operate on large matrices involved 
in the attention operation, which limits hardware utilization. In order to address these challenges, this work proposes a 
novel dynamic inference scheme, DynaTran, which prunes activations at runtime with low overhead, substantially reducing 
the number of ineffectual operations. This improves the throughput of transformer inference. We further propose tiling the matrices in transformer operations along with diverse dataflows to improve data reuse, thus enabling higher energy
efficiency. To effectively implement these methods, we propose AccelTran, a novel accelerator architecture for 
transformers. Extensive experiments with different models and benchmarks demonstrate that DynaTran achieves higher
accuracy than the state-of-the-art top-$k$ hardware-aware pruning strategy while attaining up to 1.2$\times$ higher
sparsity. One of our proposed accelerators, AccelTran-Edge, achieves 330K$\times$ higher throughput with 93K$\times$ 
lower energy requirement when compared to a Raspberry Pi device. On the other hand, AccelTran-Server achieves 
5.73$\times$ higher throughput and 3.69$\times$ lower energy consumption compared to the state-of-the-art transformer 
co-processor, Energon. \textcolor{black}{The simulation source code is available at \url{https://github.com/jha-lab/acceltran}.}

\end{abstract}


%
\IEEEpeerreviewmaketitle

\section{Introduction}

\IEEEPARstart{T}{he} transformer architecture~\cite{bert}, which is based on the self-attention 
mechanism~\cite{vaswani}, has gained widespread interest in the domain of natural language 
processing~\cite{efficient_txf_survey} and, recently, even in computer vision~\cite{vit_2021}. One reason is its 
massive parallelization capabilities on modern-day graphical processing units (GPUs), unlike traditional 
sequential models like long short-term memories~\cite{lstm} and recurrent neural networks~\cite{rnn} 
that are slow to train and thus may not perform as well. Transformers have been able to achieve state-of-the-art 
performance on diverse benchmarking tasks due to pre-training on massive public and private language 
corpora~\cite{roberta, turing_nlg, evolved_txf}.

The massive models come with their own challenges. For instance, pre-training a large 
state-of-the-art model usually requires millions of dollars worth of GPU resources~\cite{gpt_3}. Furthermore, large 
transformer models also have a high memory footprint, making them challenging to train even on modern GPUs. 
Convolutional neural networks (CNNs) have been able to overcome these challenges with a plethora of 
application-specific integrated circuit (ASIC)-based accelerators, each specialized for a different set
of models in its design space~\cite{eyeriss, spring}. These accelerators have specially-designed hardware 
modules that leverage sparsity in model weights, data reuse, optimized dataflows, and CNN mapping to attain 
high performance and energy efficiency~\cite{VivienneSzeBook}. However, CNN accelerators
are incompatible with transformer workflows since they are optimized for the inner-product operation, 
the basis of a convolution operation, and not for matrix-matrix multiplication control flows.

Some recent works attempt to accelerate transformers by reducing their memory footprint and 
the compute overhead of the self-attention operation. For instance, A$^3$~\cite{a3} contains several 
approximation strategies to avoid computing attention scores close to zero. SpAtten~\cite{spatten} 
leverages a cascade token pruning mechanism that progressively prunes unimportant tokens based on low 
attention probabilities, reducing overall compute complexity. However, the proposed 
`top-$k$' pruning mechanism~\cite{spatten}, a state-of-the-art hardware-aware dynamic inference method, has 
a high compute overhead, which partially offsets its throughput gains during model inference according to our experiments (details in Section~\ref{sec:results_dynatran}). 
Energon~\cite{energon} approximates the top-$k$ pruning method with its mixed-precision multi-round 
filtering algorithm. However, it only exploits sparsity in the attention probabilities, not in 
all possible multiplication operations in the transformer architecture (details in 
Section~\ref{sec:sparsity_sa}). To tackle this problem, OPTIMUS~\cite{optimus} uses a set-associative 
rearranged compressed sparse column (SA-RCSC) format to eliminate ineffectual multiply-and-accumulate 
(MAC) operations. However, it only exploits sparsity in the weight matrices and not the activations, i.e., 
the matrices formed from intermediate MAC operations. It also only works with encoder-decoder models, where 
the decoder is known to support limited parallelism. Leveraging encoder-only models, which have recently shown to perform well even on translation and language generation tasks~\cite{bert_nmt, bert_nlg}, not 
only reduces the critical path by 2$\times$ but also improves hardware utilization. Further, these works 
implement an entire matrix multiplication over an array of processing elements (PEs), which are the basic 
compute blocks of an accelerator. \textcolor{black}{OPTIMUS~\cite{optimus}, with its SA-RCSC sparse matrix format, does not 
break down the matrices involved into multiple \emph{tiles} \textcolor{black}{[implemented in general matrix multiplication (GEMM) pipelines]} in order to improve hardware utilization. FTRANS~\cite{ftrans} 
and SpAtten~\cite{spatten} break down a matrix-matrix multiplication operation into multiple matrix-vector multiplication 
operations, losing out on data reuse capabilities. This also limits the scope of parallelization (details in 
Section~\ref{sec:results_dynatran}). Data reuse, parallelization, and optimal hardware utilization are crucial to obtaining 
high throughput and energy efficiency.} Energon~\cite{energon} is a co-processor and not a full-fledged accelerator. This limits the 
scope of optimization across the entire pipeline, resulting in superfluous off-chip accesses. 
Field-programmable gate array (FPGA)-based transformer accelerators have also been proposed owing to their 
low cost~\cite{ftrans, fpga_accelerator_1, fpga_accelerator_2}. However, they suffer from performance 
and power inefficiencies due to bit-level reconfigurable abstractions and correspondingly high interconnect 
overheads~\cite{plasticine}.

To overcome the above challenges, we propose AccelTran, a novel cycle-accurate accelerator for 
transformer models. Our main contributions are as follows.

\begin{itemize}
    \item We propose a granular and hardware-aware dynamic inference framework, DynaTran, for transformers that dynamically 
prunes all activations in order to remove ineffectual MAC operations. DynaTran has much less compute overhead compared to 
previous works~\cite{spatten, energon}, enabling higher throughput for model inference. 
    \item To \emph{efficiently} execute DynaTran, we design and implement an ASIC-based architecture
called AccelTran. Instead of using traditional encoder-decoder models~\cite{optimus}, we leverage 
recently-proposed encoder-only models~\cite{bert}, thus reducing the critical path by 2$\times$ and improving throughput and 
hardware utilization. \textcolor{black}{Further, unlike previous works~\cite{energon}, AccelTran's 
dynamic inference pipeline is agnostic to the pre-processed weight pruning strategy.}
    \item We propose the use of \emph{tiled} matrix multiplication for our transformer accelerator. For this, 
we leverage a novel mapping scheme from the transformer model to the tiled operations that maximizes hardware 
utilization and improves parallelization.
    \item We also formulate and implement, for the first time, various \emph{dataflows} for the transformer 
optimal dataflow that maximizes data reuse to improve energy efficiency.
    \item We further leverage monolithic-3D RRAM~\cite{monolithic_3d_rram} for higher memory bandwidth. This 
alleviates the performance bottleneck in transformer inference since state-of-the-art models are huge and thus
memory-bound~\cite{turing_nlg, hat_mit}. \textcolor{black}{Our proposed control block maps the transformer computational
graph to scheduled hardware-implementable operations. It leverages the high-bandwidth monolithic-3D RRAM to schedule
these operations intelligently, enabling high throughput and energy efficiency. We also support LP-DDR3 memory for low-cost edge solutions.}
\end{itemize}

The rest of the article is organized as follows. Section~\ref{sec:background} presents background on 
transformer acceleration. Section~\ref{sec:methodology} illustrates the methodology underpinning the DynaTran and 
AccelTran frameworks in detail. Section~\ref{sec:exp_setup} describes the experimental setup and baselines 
that we compare against. Section~\ref{sec:results} discusses the results. Section~\ref{sec:discussion} 
compares related works and suggests future work directions. Finally, Section~\ref{sec:conclusion} concludes 
the article.

\section{Background and Motivation}
\label{sec:background}

In this section, we provide background on various compute operations employed in a transformer model and 
previous works on transformer pruning and dynamic inference (sometimes interchangeably termed as dynamic 
pruning~\cite{spatten, energon}).

\subsection{The Transformer Model}
\label{sec:background_txf_model}

We present the details of the memory and compute operations in the transformer model next.

\subsubsection{Compute Operations}

\begin{table}[]
\caption{Memory and compute operations in a transformer.}
\centering
\begin{tabular}{@{\hskip 0.2in}l@{\hskip 0.5in}l@{\hskip 0.2in}}
\toprule
\multicolumn{2}{c}{Word Embedding and Position Encoding}                                                                                      \\ \midrule
\textbf{M-OP-0}         & $\mathbf{H} = \mathbf{H}_{emb} + \text{PE}(\mathbf{H}_{emb})$                                                                \\ \midrule
\multicolumn{2}{c}{Multi-Head Attention}                                                                                                      \\ \midrule
\textbf{M-OP-{[}1-4{]}} & load $\mathbf{W}^\text{Q}_i$, $\mathbf{W}^\text{K}_i$, $\mathbf{W}^\text{V}_i$, $\mathbf{W}^\text{O}_i$                                                  \\
\textcolor{mac}{\textbf{C-OP-{[}1-3{]}}} & $\mathbf{Q}_i, \mathbf{K}_i, \mathbf{V}_i = \mathbf{H} \mathbf{W}^\text{Q}_i, \mathbf{H} \mathbf{W}^\text{K}_i, \mathbf{H} \mathbf{W}^\text{V}_i$ \\
\textcolor{mac}{\textbf{C-OP-4}}         & $\mathbf{A}_i = \mathbf{Q}_i \mathbf{K}_i$                                                                         \\
\textcolor{sftm}{\textbf{C-OP-5}}         & $\mathbf{S}_i = \text{softmax}\left( \frac{\mathbf{A}_i}{\sqrt{h}} \right)$                                                  \\
\textcolor{mac}{\textbf{C-OP-6}}         & $\mathbf{P}_i = \mathbf{S}_i \mathbf{V}_i$                                                                                   \\
\textcolor{mac}{\textbf{C-OP-7}}         & $\mathbf{H}^{\text{MHA}}_i = \mathbf{P}_i \mathbf{W}^\text{O}_i$                                                                           \\ \midrule
\multicolumn{2}{c}{Add and Layer-norm}                                                                                                                \\ \midrule
\textcolor{ln}{\textbf{C-OP-8}}         & $\mathbf{H}^{\text{LN}} = \text{layer-norm}(\mathbf{H}^{\text{MHA}} + \mathbf{H})$                                                                      \\ \midrule
\multicolumn{2}{c}{Feed Forward}                                                                                                              \\ \midrule
\textbf{M-OP-{[}5-6{]}} & load $\mathbf{W}^{\text{F1}}, \mathbf{W}^{\text{F2}}$                                                                                      \\
\textcolor{mac}{\textbf{C-OP-9}}         & $\mathbf{H}^{\text{F1}} = \text{GeLU}(\mathbf{W}^{\text{F1}} \mathbf{H}^{\text{LN}})$                                                             \\
\textcolor{mac}{\textbf{C-OP-10}}        & $\mathbf{H}^{\text{F2}} = \text{GeLU}(\mathbf{W}^{\text{F2}} \mathbf{H}^{\text{F1}})$                                                             \\ \midrule
\multicolumn{2}{c}{Layer-norm}                                                                                                                \\ \midrule
\textcolor{ln}{\textbf{C-OP-11}}        & $\mathbf{H}^\text{O} = \text{layer-norm}(\mathbf{H}^{\text{F2}})$                                                                            \\ \bottomrule
\end{tabular}
\label{tbl:txf_ops}
\end{table}

Table~\ref{tbl:txf_ops} summarizes the required memory load and compute operations in a transformer model. The first is 
the loading of word embeddings and position encodings, which take up a significant fraction of the weights in a 
transformer. Here, $\mathbf{H}_{emb}$ corresponds to the embeddings of all tokens in the vocabulary (vocabulary size is 
30,522 for the BERT~\cite{bert} family of models). We represent each token by a vector of length $h$, which is the 
hidden dimension of the transformer (e.g., $h=128$ for BERT-Tiny~\cite{turc2019} and $h=768$ for BERT-Base~\cite{bert}). 
Then, we load the weight matrices for the multi-head attention operations. Here, $\mathbf{W}^\text{Q}_i$, 
$\mathbf{W}^\text{K}_i$, and $\mathbf{W}^\text{V}_i \in \mathbb{R}^{h \times h/n}$ are needed in each attention head, 
where $n$ is the number of attention heads. Subsequent compute operations (color-coded \textcolor{mac}{\textbf{blue}} for matrix 
multiplication and \textcolor{sftm}{\textbf{green}} for softmax) are employed in self-attention~\cite{vaswani}. Intermediate 
matrices are called \emph{activations}; those that are loaded from memory are called \emph{weights}. $\mathbf{W}^\text{O}_i \in \mathbb{R}^{h/n \times h/n}$ maps the attention probabilities to output scores. Then, we add the 
input to the output of the multi-head attention (which is formed by concatenating the output of all attention heads) and 
normalize the resultant matrix. This is the layer-norm operation (color-coded \textcolor{ln}{\textbf{orange}}) that is used to 
reduce covariance shifts~\cite{layer_norm}. Finally, the layer norm feeds the feed-forward operation that, in turn, 
feeds the layer norm. GeLU is the activation function commonly used in transformers~\cite{bert, gelu}.

\subsubsection{Memory Requirements}

\begin{figure}
    \centering
    \includegraphics[width=\linewidth]{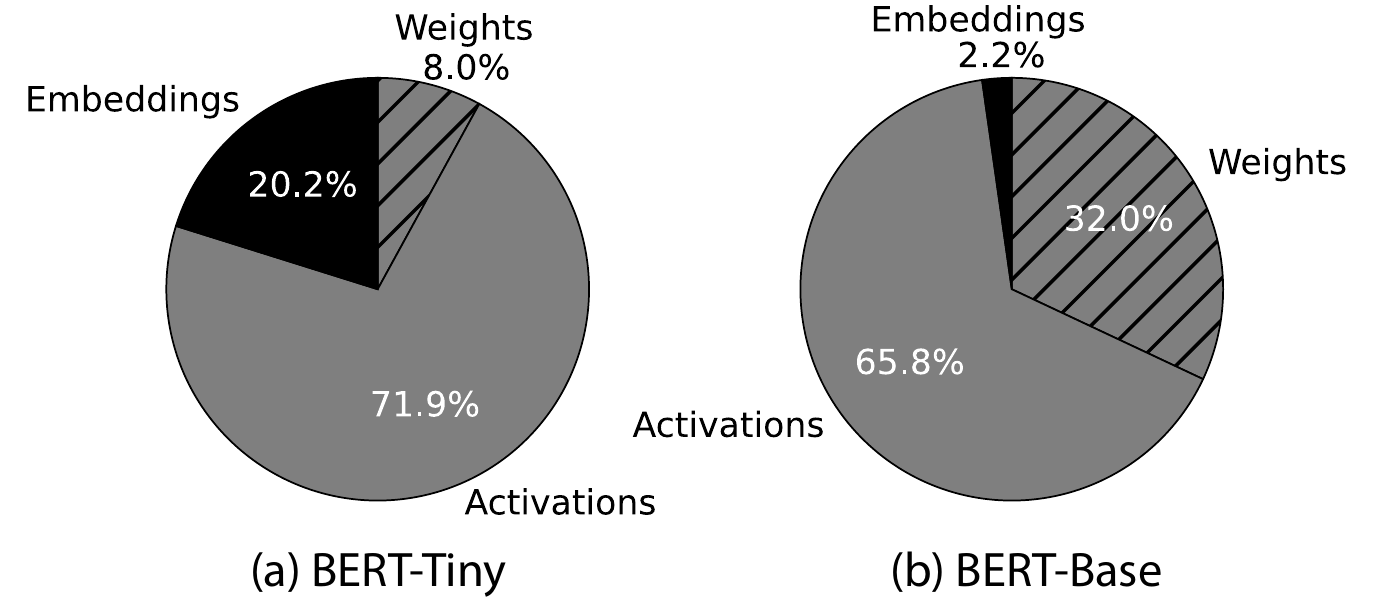}
    \caption{Memory requirements for (a) BERT-Tiny and (b) BERT-Base.}
    \label{fig:mem_req}
\end{figure}

Fig.~\ref{fig:mem_req} shows the memory requirements for BERT-Tiny and BERT-Base. BERT-Tiny has higher memory
requirements for word and position embeddings (compared with BERT-Base) relative to requirements for weights and activations. Further, activations 
take up much memory, 8.98$\times$ that of the weights for BERT-Tiny and 2.06$\times$ for BERT-Base. The total main 
memory requirements for the two models are 52.8MB and 3.4GB, respectively, when only the weights and embeddings are stored. 
Activations are formed at runtime and stored in internal registers or on-chip buffers. With increasing transformer model
sizes (calculated solely in terms of weights)~\cite{turing_nlg}, taking into account their operation on hardware 
accelerators, the memory budget should also have to account for the commensurate increase in activations.

\subsection{Sparsity in Self-Attention}
\label{sec:sparsity_sa}

Researchers have striven to reduce the computational complexity of transformers by pruning, during 
pre-training or fine-tuning, the transformer weights~\cite{compressing_bert, movement_pruning}. Previous works have also proposed various methods to reduce the quadratic complexity of the self-attention operation~\cite{linformer}. Distillation~\cite{distillbert} recovers the accuracy loss 
due to such pruning techniques. However, all these works prune the 
model while training; more so, they only prune the weights. During inference, sparse matrices with ineffectual values may be formed \emph{dynamically} from both activations and weights. Such ineffectual values must be pruned at runtime to improve energy efficiency and hardware utilization. 

SpAtten~\cite{spatten} proposed the top-$k$ pruning method. It essentially identifies query-key pairs that produce large 
attention probabilities at runtime. Given an attention score matrix ($\mathbf{S}_i$ in Table~\ref{tbl:txf_ops}), it keeps 
the $k$ largest elements in each row to obtain the probability matrix ($\mathbf{P}_i$) and neglects the rest. Even though 
this method only results in a minor accuracy loss, it has a high overhead (as we show experimentally in 
Section~\ref{sec:results_dynatran}) due to its $\mathcal{O}(N^3)$ complexity. Further, a matrix multiplication operation 
benefits from sparsification when small values, which do not have much effect on the final result, are completely pruned 
out so that the hardware does not have to implement the corresponding MAC operations. SpAtten only considers the attention 
probabilities ($\mathbf{P}_i$), but not all the matrix multiplication operations presented in Table~\ref{tbl:txf_ops}. Thus, it
loses out on gains that could be obtained by pruning other matrices as well. We compare it with our proposed method, 
DynaTran, in Section~\ref{sec:results_dynatran}.

\section{Methodology}
\label{sec:methodology}

\begin{figure}
    \centering
    \includegraphics[width=\linewidth]{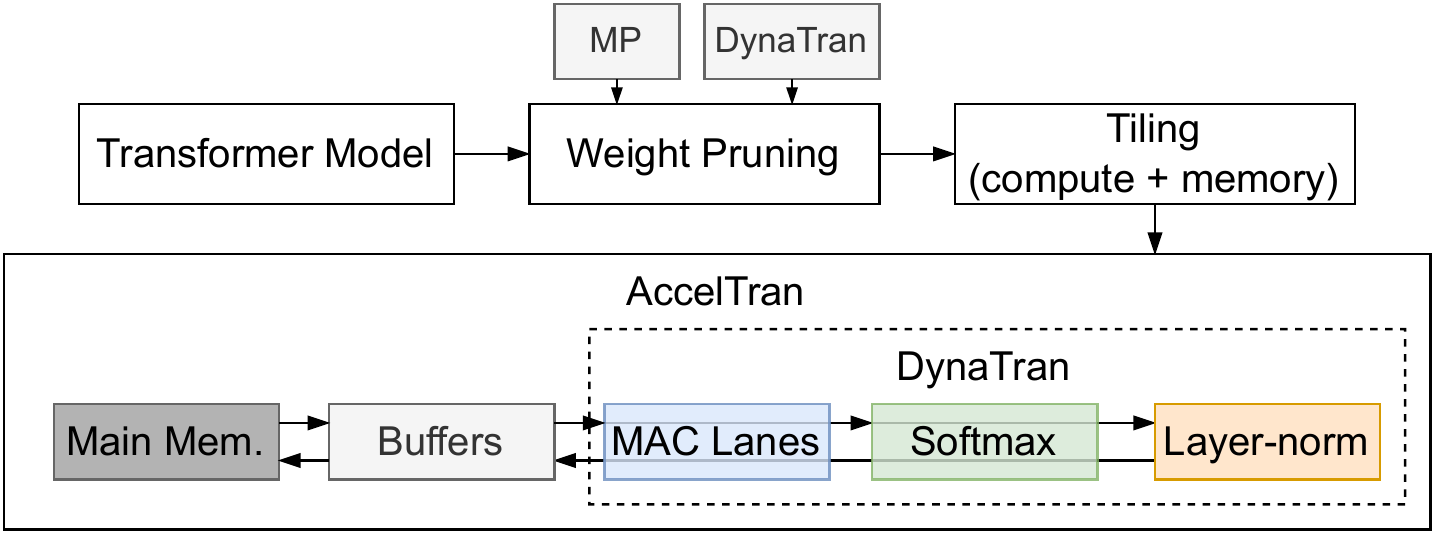}
    \caption{AccelTran workflow for an input transformer model and its acceleration in hardware.}
    \label{fig:flowchart}
\end{figure}

Fig.~\ref{fig:flowchart} presents a flowchart for the AccelTran simulation pipeline. We first weight-prune 
the transformer that is provided as input, either using movement pruning (MP)~\cite{movement_pruning} or DynaTran. 
Then, we tile the transformer model into granular compute and memory operations. These tiled operations 
are passed to the AccelTran simulator, which implements the tiled operations, in hardware, in a 
cycle-accurate manner.

We now present the DynaTran framework for efficient dynamic inference with the transformer model. We also present AccelTran, a cycle-accurate accelerator for implementing this framework efficiently in hardware.

\subsection{DynaTran}
\label{sec:dynatran}

Unlike the top-$k$ pruning algorithm~\cite{spatten}, we propose a low-overhead dynamic inference method that quickly prunes ineffectual weight and activation values at runtime. For a given matrix, which 
is either loaded as a weight matrix from memory or is an activation matrix obtained from previous MAC 
operations, DynaTran prunes values with a magnitude less than a given threshold $\tau$. Mathematically, 
an input matrix $\mathbf{M} \in \mathbb{R}^{m \times n}$ is pruned to $\mathbf{M}^\text{P}$ as follows:
\begin{equation*}
    \mathbf{M}^\text{P}_{ij} = 
    \begin{cases*} 
        \mathbf{M}_{ij} & if $ \lvert \mathbf{M}_{ij} \rvert \ge \tau $  \\
        0 & if $ \lvert \mathbf{M}_{ij} \rvert < \tau $
    \end{cases*}
\end{equation*}

This simple comparison operation incurs negligible compute overhead at runtime. This is important since 
transformer evaluation involves many 
such matrices at runtime, most of which are on the critical path for model 
computation. Further, each comparison operation can be parallelized, ensuring that pruning only takes 
up one clock cycle. This has a much lower overhead compared to SpAtten~\cite{spatten} and 
Energon~\cite{energon} that have dedicated engines for this operation. We now define the pruning ratio 
(or level of sparsity) for the output matrix as:
\begin{equation*}
    \rho(\mathbf{M}^\text{P}) = \frac{\sum_{x \in \mathbf{M}^\text{P}} \delta_{x, 0}}{m \times n}
\end{equation*} where $\delta$ is the Kronecker delta function. We profile the resultant sparsity in the weights and activations for different transformer models on diverse applications to obtain a desired $\rho$. One or more such profiled curves can be stored in memory. For the desired values of $\rho$, we determine the corresponding $\tau$ at runtime through a simple look-up operation. We present such curves in 
Section~\ref{sec:results_dynatran} to compare the throughput of our proposed approach with top-$k$ pruning.

\subsection{The AccelTran Simulator}

We present details of the proposed accelerator simulator next.

\subsubsection{Tiling and Dataflow}

\begin{figure}
    \centering
    \includegraphics[width=\linewidth]{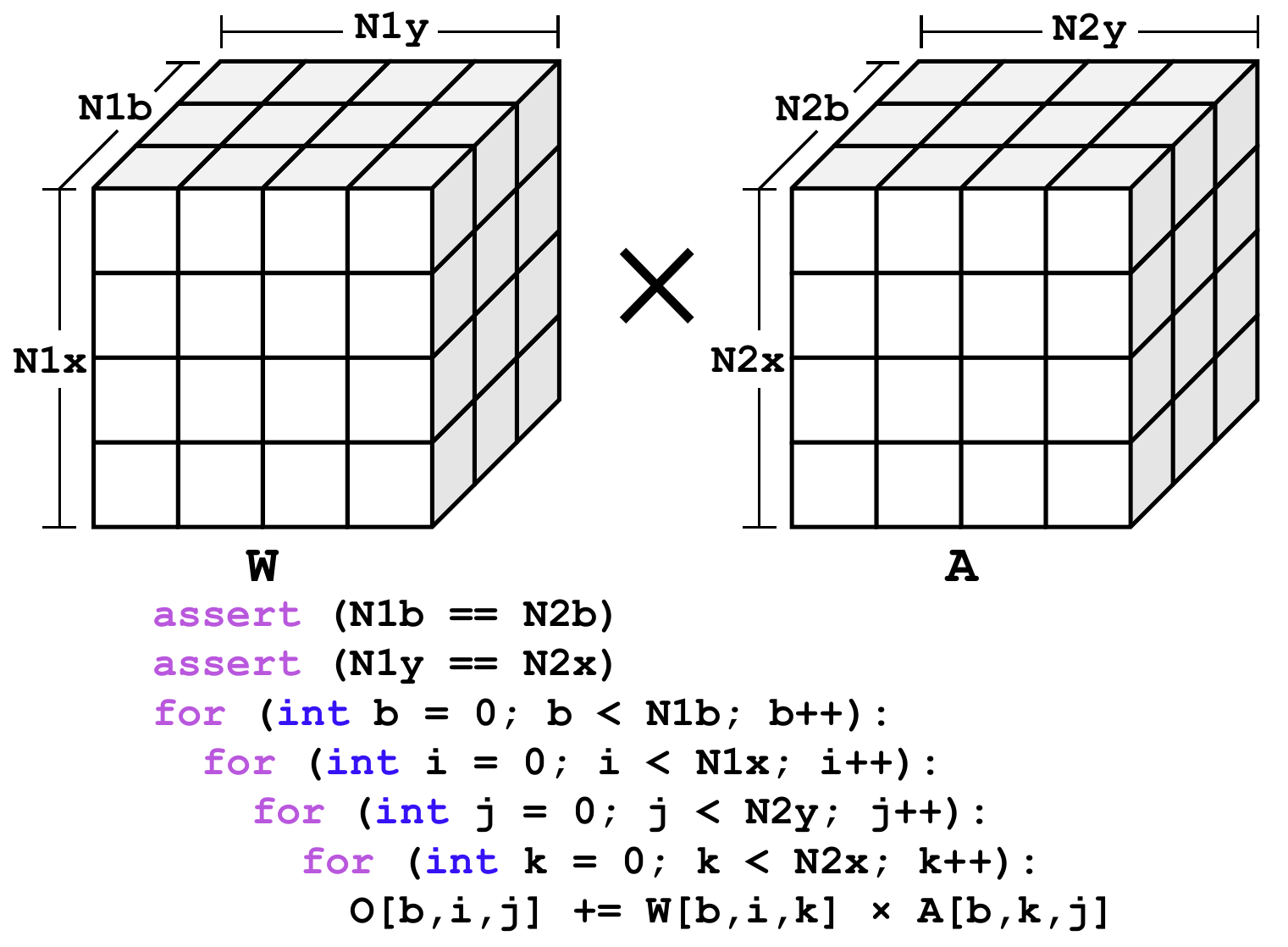}
    \caption{Tiling of a matrix multiplication operation along with a selected dataflow (specifically,
\code{[b,i,j,k]}). Here, a tensor is shown instead, with the first dimension
being the batch size.}
    \label{fig:tiling_dataflow}
\end{figure}

As per Table~\ref{tbl:txf_ops}, most compute operations in the transformer model are matrix multiplication 
operations. Thus, it is important to optimize these operations for high gains. Unlike previous works
that perform matrix multiplications directly using large MAC units, we propose using tiled matrix 
multiplication (primarily employed by modern GPUs~\cite{tiling_gpu}). Tiling the operations helps 
with better utilization of resources and enables massive parallelization. Fig.~\ref{fig:tiling_dataflow} 
shows the tiling operation along with an example \emph{dataflow}. We can also think of a dataflow as a 
loop-unrolling scheme. The four for-loops can be unrolled in any permutation (giving 24 possible ways to 
unroll the loops, i.e., 24 dataflows). Multiplication between two tiles (say, weights \code{W[b,i,k]} and activations \code{A[b,k,j]}) is performed by a MAC lane (in parallel, based on the number of MAC units).

Each dataflow results in different data reuse capabilities. For example, if only four MAC lanes are 
available, with the dataflow shown in Fig.~\ref{fig:tiling_dataflow}, when \code{j} changes from 
\code{0} to \code{1} (\code{b} and \code{i} remaining constant), the MAC lanes can reuse the corresponding weights
\code{W[b,i,k]}, \code{k} $\in$ \code{[0,...,N2x]}. Similarly, other dataflows would result in different 
reuse capabilities for different input matrix sizes. We show the reuse instances and corresponding energy 
savings for this example in Section~\ref{sec:results_dataflow}. \textcolor{black}{No previous work has leveraged different dataflows to improve data reuse in transformer evaluation.}

\subsubsection{Accelerator Organization}

Taking inspiration from a state-of-the-art CNN accelerator, SPRING~\cite{spring}, we leverage monolithic-3D 
integration to connect to an on-chip 3D resistive random-access memory (RRAM)~\cite{monolithic_3d_rram}. In monolithic-3D 
integration, multiple device tiers are fabricated on one substrate wafer, connected through monolithic 
inter-tier vias that allow much higher density than traditional through-silicon-via-based 3D 
integration~\cite{miv}. This leaves much more space for logic and also permits high memory bandwidth, which
are crucial for large state-of-the-art transformer models. For scalable edge deployments, we also support an off-chip dynamic RAM (DRAM).

\begin{figure}[t]
    \centering
    \includegraphics[width=0.85\linewidth]{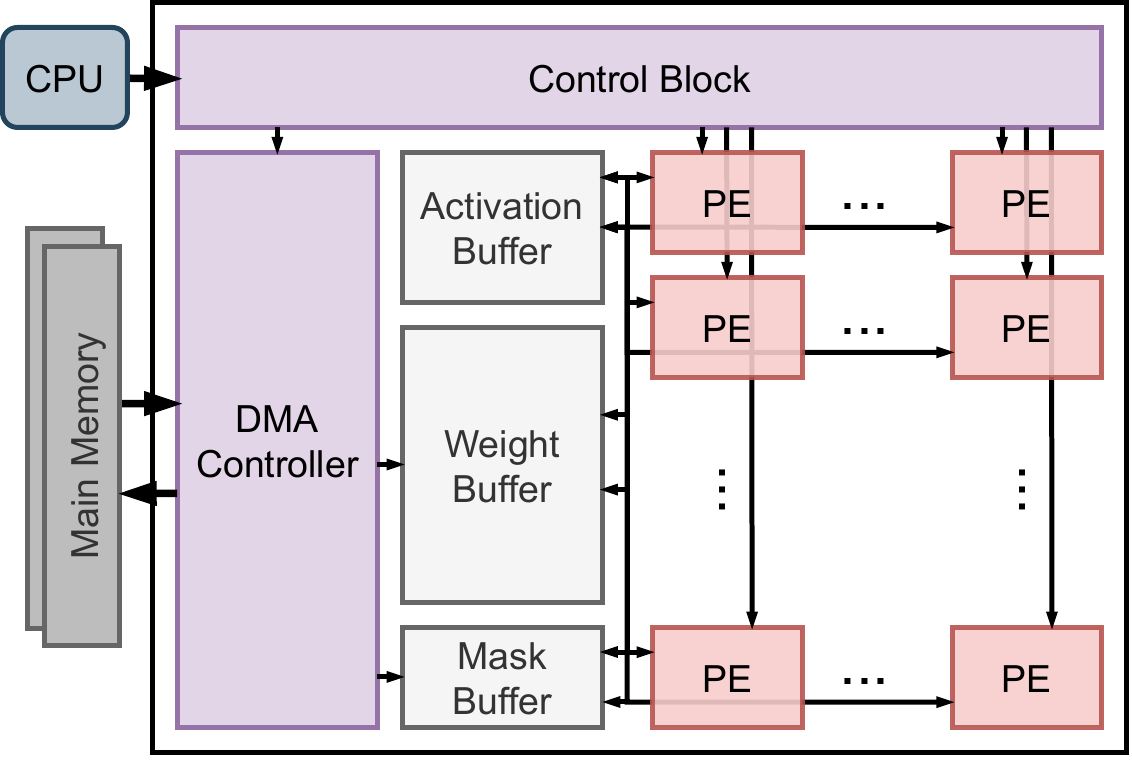}
    \caption{Accelerator organization.}
    \label{fig:acceltran_org}
\end{figure}

Fig.~\ref{fig:acceltran_org} shows the organization of the accelerator tier in the proposed architecture. The 
control block takes the instruction stream for the transformer model from the host CPU. The weights and 
embeddings are brought on-chip from the off-chip DRAM, or from the monolithic-3D RRAM, by the direct memory access (DMA) controller. The activation and the weight buffers store the activations and weights, respectively, in a compressed format (discussed 
in Section~\ref{sec:sparsity_aware}). Data compression relies on binary masks (stored in the mask buffer). 
The PEs use the compressed data and the associated masks to perform the main compute operations in the transformer.

\subsubsection{Processing Elements}

\begin{figure}[t]
    \centering
    \includegraphics[width=0.85\linewidth]{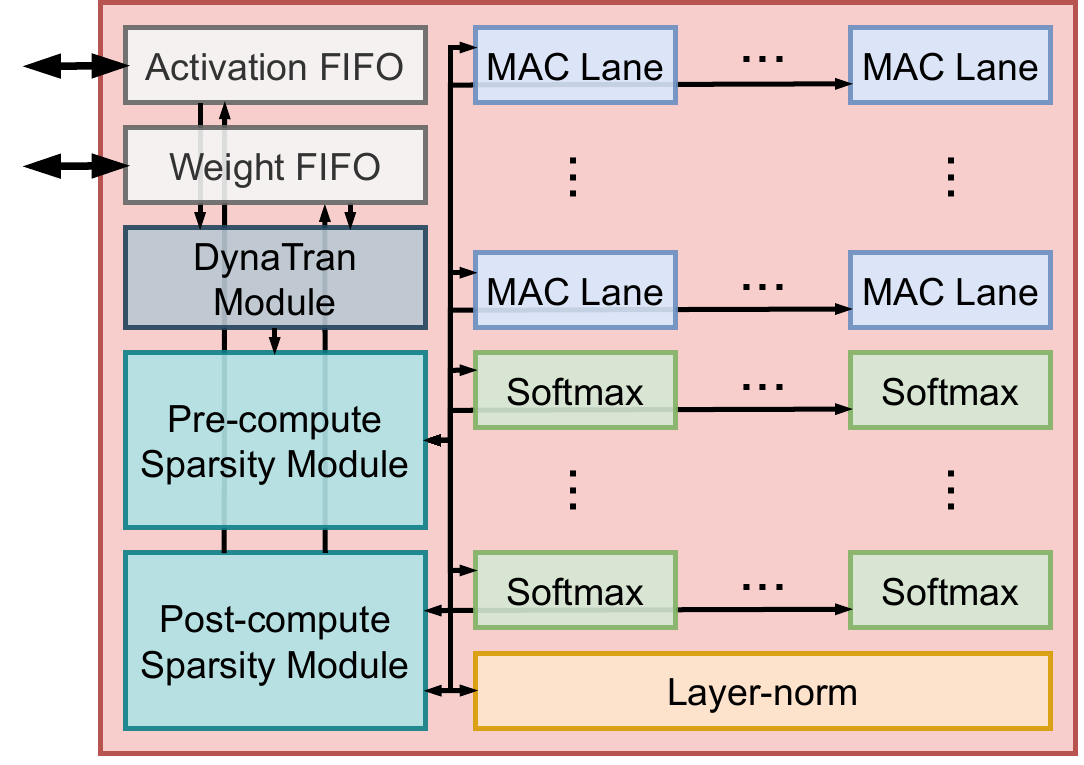}
    \caption{Internal components of a PE.}
    \label{fig:pe}
\end{figure}

Fig.~\ref{fig:pe} shows the main modules present inside a PE, which is the basic compute block in our accelerator. 
The compressed data are stored in local registers of the PE by the activation first-in-first-out (FIFO) and 
weight FIFO registers. The data then enter the DynaTran module that induces sparsity based on the desired $\rho$. 
As explained in Section~\ref{sec:dynatran}, this module prunes the given weights or activations based on a 
pre-calculated threshold $\tau$. The sparse data then enter the pre-compute sparsity module with the binary masks. 
This module converts the input data into a zero-free format based on the associated masks. The PE then forwards this zero-free data to the MAC lanes (for matrix multiplication), softmax modules (for softmax operation), or the 
layer-norm module (for layer-norm operation). The zero-free data eliminate any ineffectual computations in these 
modules. Finally, the post-compute sparsity module implements the inverse of this operation on the output activations, before storing them 
in the activation FIFO register and, eventually, the main activation buffer.

\subsubsection{MAC Lanes}

\begin{figure}[!t]
    \centering
    \includegraphics[width=0.9\linewidth]{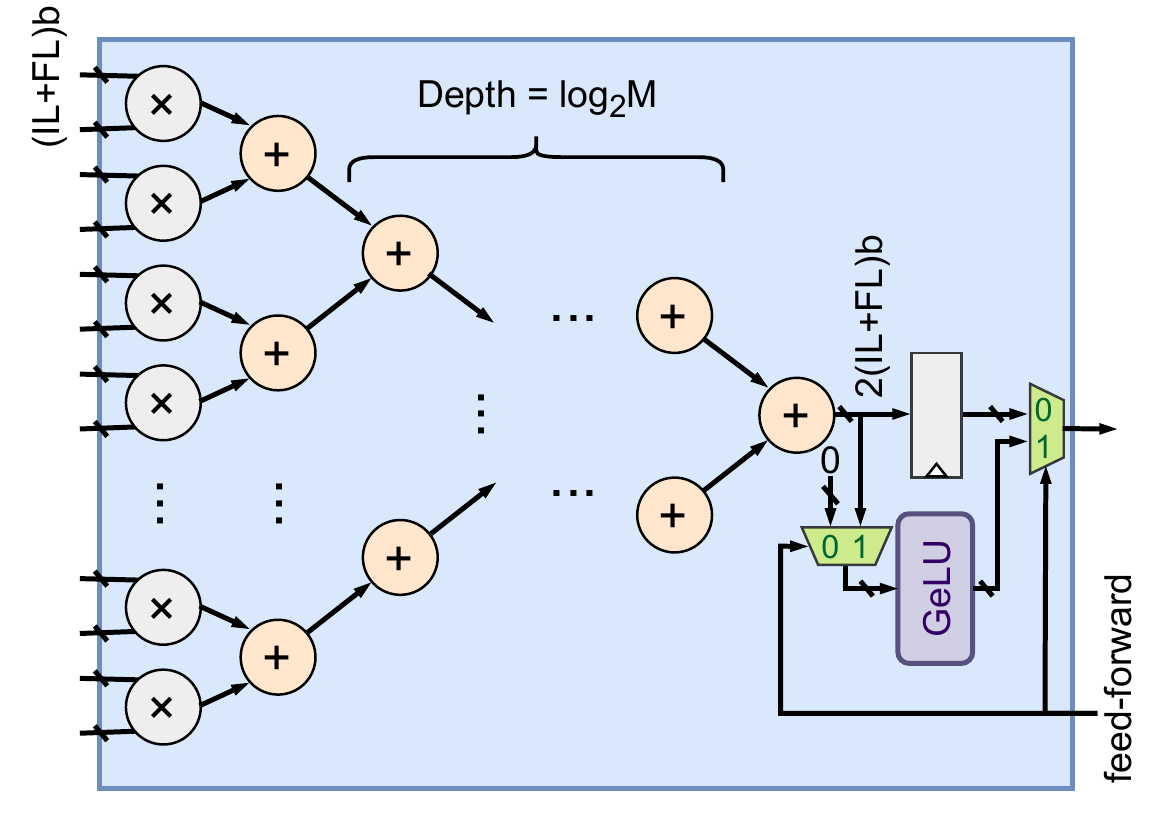}
    \caption{Architecture of the MAC Lane.}
    \label{fig:mac_lane}
\end{figure}

MAC lanes are responsible for multiplication between two tiles in a parallelized fashion. Let the tiles be 
denoted by $\mathbf{W} \in \mathbb{R}^{b \times x \times y}$ and $\mathbf{A} \in \mathbb{R}^{b \times y \times z}$ 
for conserved matrix (in general, tensor) multiplication. Then, the number of multiplication operations is 
$n_o = b \times x \times y \times z$. Each MAC lane in AccelTran has $M$ multipliers. Thus, the minimum number 
of cycles to compute the tiled operation is $n_o / M$. Fig.~\ref{fig:mac_lane} shows the implementation 
of a MAC lane. We store all activation and weight data in fixed-point format with $(\text{IL} + \text{FL})$ 
bits, denoting integer length and fractional length, respectively~\cite{spring}. 
The module first feeds the data to the $M$ multipliers, then the corresponding outputs to the adder tree over multiple stages. We represent the products with $2 \times (\text{IL} + \text{FL})$ 
bits to prevent overflow. The accumulations also use this bit-width. The depth of the adder tree is
$\log_2 M$ for the $M$ multipliers in our MAC lane. The module then passes the data to the output register. For 
feed-forward operations, where activation is required, the GeLU module implements this nonlinearity at the 
output of the MAC units. All other compute modules also work with the $(\text{IL} + \text{FL})$ bits. 

\subsubsection{Dynamic Inference Modules}
\label{sec:dynatran_module}

\begin{figure}
    \centering
    \includegraphics[width=0.9\linewidth]{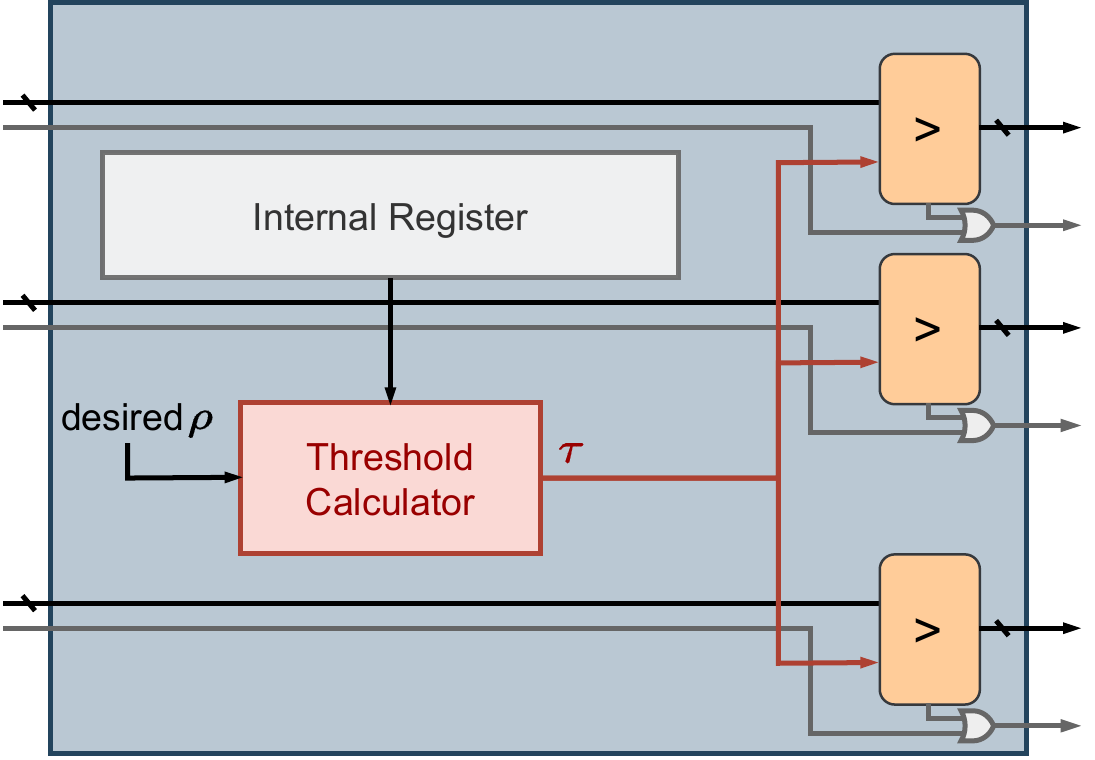}
    \caption{DynaTran module. The wires for mask bits are in grey.}
    \label{fig:dynatran_module}
\end{figure}

To execute DynaTran pruning, we implement a low-overhead DynaTran module that prunes ineffectual values in the 
input activations or weights. As explained in Section~\ref{sec:dynatran}, we prune the values of the input matrices by comparing their magnitude with a pre-determined threshold $\tau$. Fig.~\ref{fig:dynatran_module} 
shows how this is implemented, in parallel, for the entire tile. For an input tile 
$\mathbf{M} \in \mathbb{R}^{b \times x \times y}$, we use $b \times x \times y$ comparators. The threshold calculator determines the required threshold, using the desired $\rho$ and the pre-profiled transfer functions for different transformer models on diverse applications. The internal register stores these transfer functions loaded from memory before running transformer evaluation. If the output of the comparator 
is zero, we set the corresponding mask bit to one. Here, we represent the lines carrying mask information
in grey and those carrying activation/weight information in black.

\subsubsection{Sparsity-aware Acceleration}
\label{sec:sparsity_aware}

\begin{figure}
    \centering
    \includegraphics[width=\linewidth]{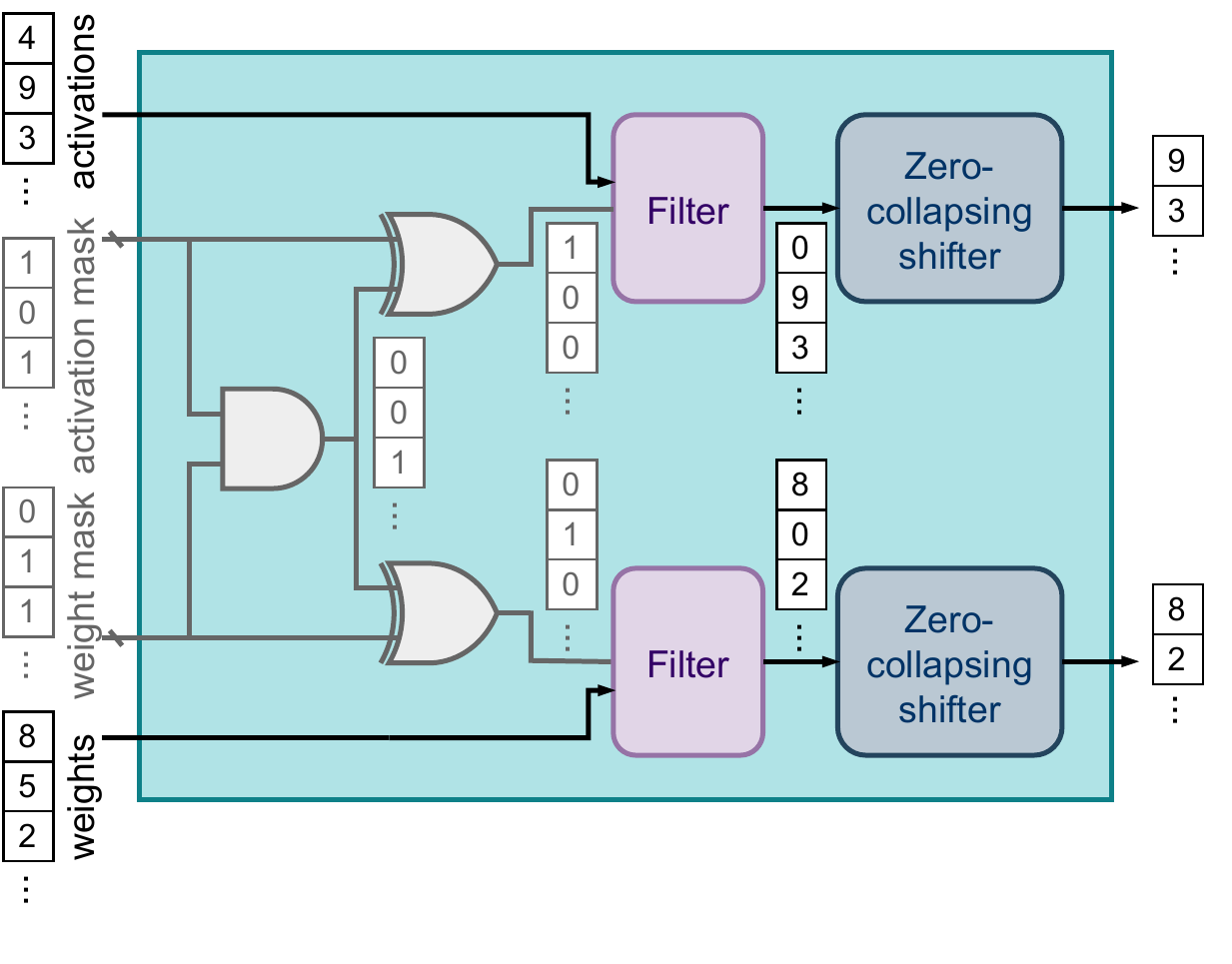}
    \caption{Pre-compute sparsity module.}
    \label{fig:pre_sparsity}
\end{figure}

To exploit sparsity and skip ineffectual activations and weights, and reduce memory footprint, AccelTran uses a 
binary-mask scheme to encode the sparse data and perform computations directly in the encoded format. Compared 
to the regular dense format, the pre-compute sparsity module compresses data by removing all the zero elements. 
In order to retain the shape of the uncompressed data, we use an extra binary mask~\cite{spring}. The binary 
mask has the same shape as the uncompressed data, where each binary bit in the mask is associated with one 
element in the original data vector. If the entry in the mask is 1, it means that the corresponding 
activation/weight entry is ineffectual and should not be used for further computation. 

Fig.~\ref{fig:pre_sparsity} illustrates the pre-compute sparsity module. It takes the zero-free data 
and binary mask vectors as inputs and generates an output mask and zero-free activations/weights for
the MAC lanes, softmax modules, or the layer-norm module. The output binary mask indicates the common
indices of non-zero elements in both the activation and weight vectors. The module computes this mask using a bit-wise
\textbf{AND} function over the input activation and weight masks. The two \textbf{XOR}
gates then generate the filter masks. Based on the filter masks, the filter prunes the activations/weights. Finally, the zero-collapsing shifter compresses the activations/weights to feed zero-free data to the 
compute modules for further computation~\cite{spring}. Thus, we completely skip ineffectual computations, improving throughput and energy efficiency.

\subsubsection{Simulator Flow}

\begin{figure}
    \centering
    \includegraphics[width=\linewidth]{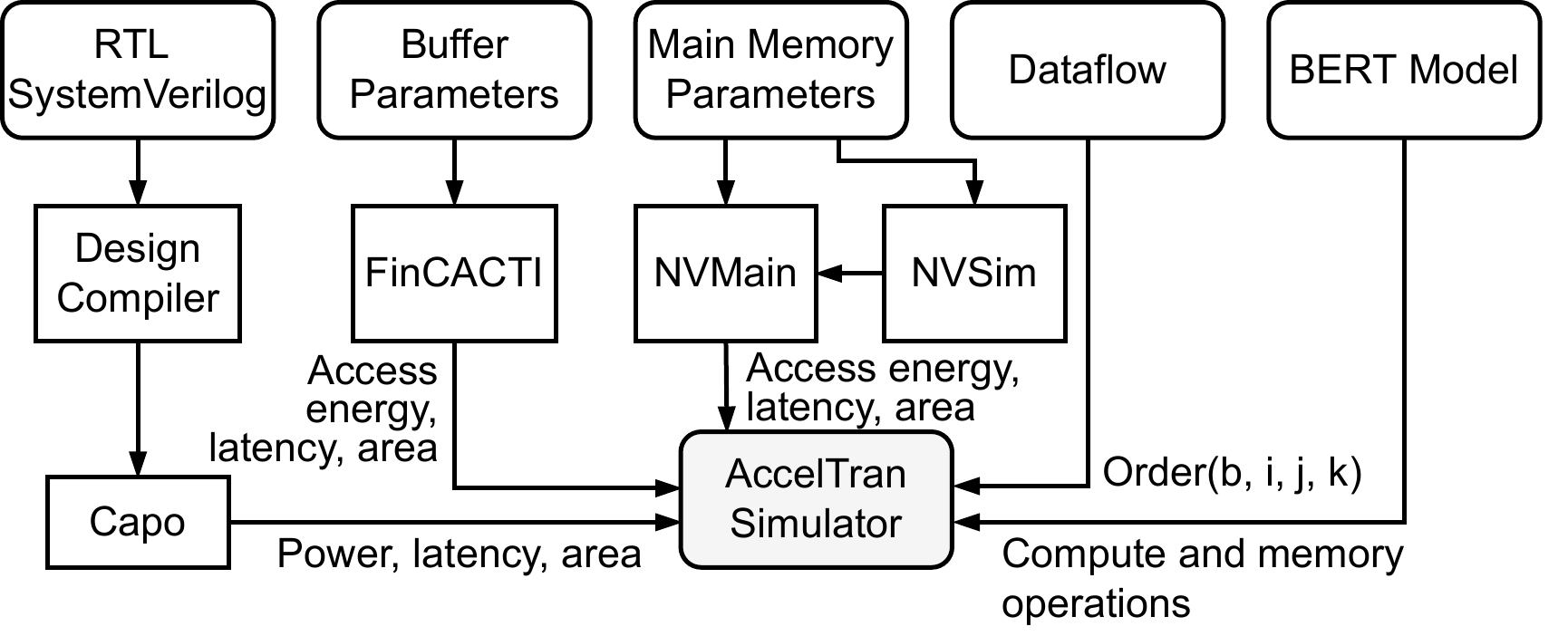}
    \caption{Flow of simulation in AccelTran.}
    \label{fig:simulator}
\end{figure}

Fig.~\ref{fig:simulator} shows the simulation flow for evaluating the AccelTran architecture. We
implement different modules presented above at the register-transfer level (RTL) with SystemVerilog. Design Compiler~\cite{dc} synthesizes the 
RTL design using a 14nm FinFET technology library~\cite{14nm}. Capo~\cite{capo}, an open-source floorplacer, performs
floorplanning. We did part of the floorplanning by hand. The net area reported is after floorplanning (including whitespaces). FinCACTI~\cite{fincacti}, a cache modeling tool for deeply-scaled FinFETs, models the on-chip buffers. NVSim~\cite{nvsim} and NVMain~\cite{nvmain} model the main memory. We then plug the synthesized results into a Python-based 
cycle-accurate simulator.

\subsubsection{Smart Scheduling of Tiled Operations}
\label{sec:smart_scheduling}

AccelTran simulates various operations in the transformer model in a tiled fashion. As discussed earlier, we tile each compute operation's activation/weight
matrices. We then assign each such tiled operation 
to a designated module based on the type of compute operation. Modules that are not being used are power-gated to reduce leakage power draw. Transformer inference may run into 
either memory or compute stalls if the corresponding prerequisites are not met. As the names suggest, a memory 
stall halts a memory operation from being executed. Similarly, a compute stall halts a compute operation. There 
is a memory stall if the buffer is not ready to load/store more data as some data are already being written or 
read. Compute operations require some activations/weights in the buffers. There could be a compute stall if 
the required matrix is not yet loaded into the buffer. A memory stall can also occur if the compute modules are using current data in the 
buffer and there is no space left to add more data. This is true until the 
current data (that are required until compute operations finish) are evicted when the corresponding compute 
operations are done and the data are no longer required. A memory stall can also occur if the compute operation is not done before storing 
activation data. Finally, if all compute modules for a specific type of compute operation are busy, it could also lead to a compute stall.

\begin{figure}
    \centering
    \includegraphics[width=\linewidth]{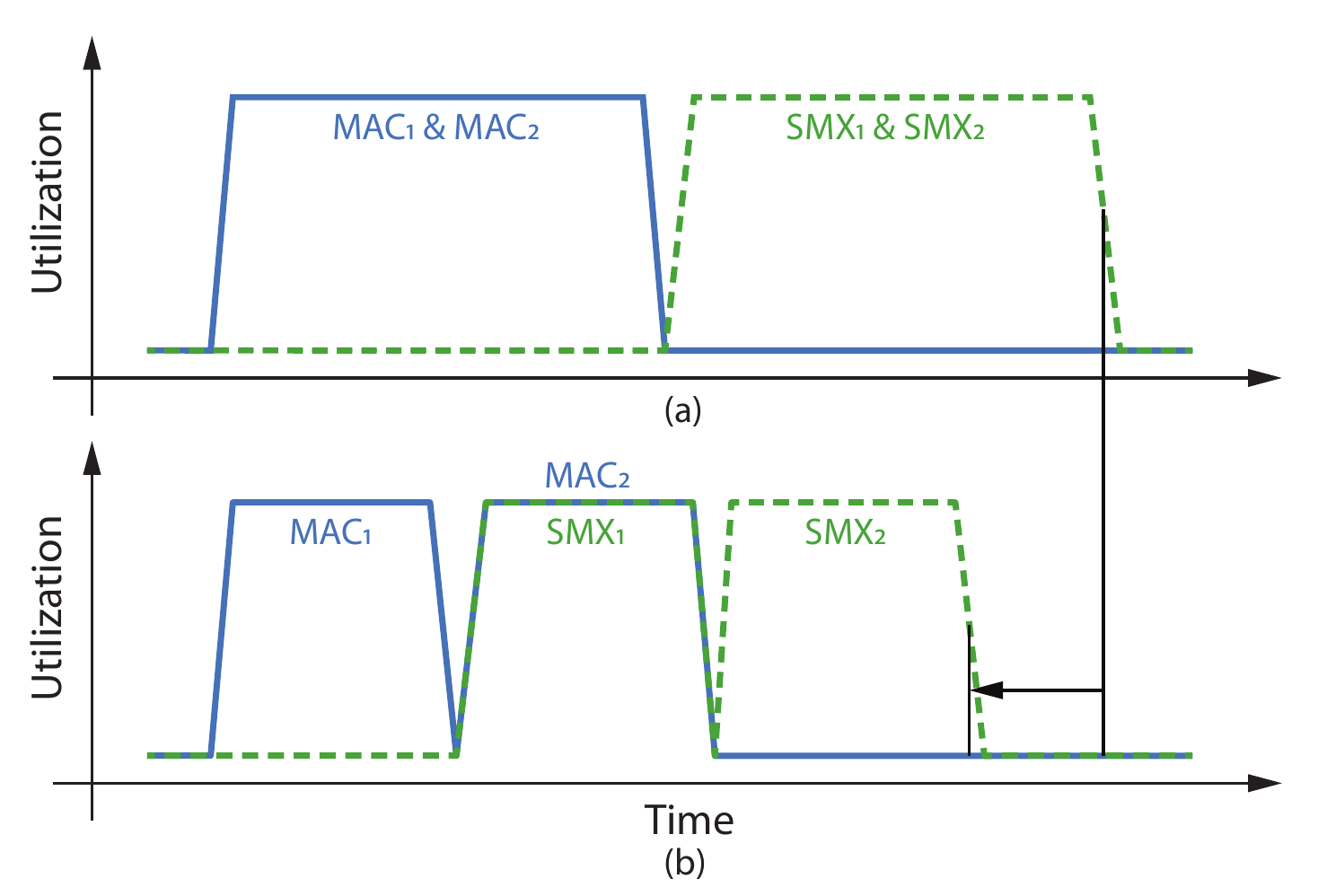}
    \caption{Scheduling with (a) equal priority and (b) staggered operations for BERT-Tiny's MAC and softmax 
(SMX) operations.}
    \label{fig:staggered_ops}
\end{figure}

The control block schedules various compute and memory operations to maximize hardware utilization. Since 
transformer models execute the same sequence of operations for every attention head, assigning equal priority 
to each head would result in poor usage of specialized resources. Hence, AccelTran staggers the operation of 
different heads. For instance, in BERT-Tiny, it gives more priority to one head so that the relevant MAC operations 
are completed first for that head. Then, when the first head reaches the softmax operation, MAC lanes can be 
assigned to the second head. This results in simultaneous utilization of the MAC lanes and softmax modules, 
thus increasing hardware utilization and improving throughput. Fig.~\ref{fig:staggered_ops} presents a working 
schematic of the staggered implementation in BERT-Tiny's MAC and softmax operations (i.e., for two attention 
heads). In the staggered case, in Fig.~\ref{fig:staggered_ops}(b), MAC lanes and softmax modules can be utilized 
simultaneously, resulting in a higher parallelization, thus leading to a higher throughput.

\section{Experimental Setup}
\label{sec:exp_setup}

In this section, we present the setup behind various experiments we performed, along with the baselines considered 
for comparison.

\subsection{Evaluation Models and Datasets}

To test the efficacy of our proposed dynamic inference method, DynaTran, we evaluate encoder-only models (because 
of their high parallelization capabilities~\cite{optimus}) on different tasks. We use BERT-Tiny~\cite{turc2019} and BERT-Base~\cite{bert}, two commonly used pre-trained models. BERT-Tiny has two encoder layers, each with a 
hidden dimension $h = 128$ and two attention heads in the multi-head attention operation, as discussed in 
Section~\ref{sec:background_txf_model}. BERT-Base is a larger model with 12 encoder layers, each with a hidden 
dimension $h = 768$ and 12 attention heads. These encoder-only models can also be extended to machine 
translation~\cite{bert_nmt} and language generation~\cite{bert_nlg}. Testing these recent extensions on hardware 
forms part of future work.

We test the two models on two \emph{representative} tasks, namely SST-2~\cite{glue} and SQuAD-v2~\cite{squad}. 
SST-2 is a popular benchmarking dataset that enables testing of model performance on sentiment analysis tasks. 
The dataset has 67K sequences in the training set and 872 in the validation set. The performance metric is the accuracy of correctly predicting label sentiment (positive or negative). SQuAD-v2 is a
popular question-answering dataset. The training and validation sets have 130K and 12K examples, respectively. The 
performance metric is the F1 score~\cite{powers2011evaluation}.

While running DynaTran, we targeted both activation and weight sparsity. Weight sparsity is static and depends 
on pruning performed during model pre-training or fine-tuning (or even DynaTran's weight pruning, as
described in Section~\ref{sec:results_dynatran_wp}). Activation sparsity changes for every input sequence and 
is reported as the average over the entire validation set.

\subsection{The AccelTran Architectures}

\begin{table}[]
\caption{Design choices for AccelTran-Edge and AccelTran-Server.}
\resizebox{\columnwidth}{!}{
\begin{tabular}{@{}c|l|l@{}}
\toprule
\textbf{Accelerator}                       & \textbf{Module}                  & \textbf{Configuration}                                     \\ \midrule
\multirow{7}{*}{\rotatebox[origin=c]{90}{\textbf{AccelTran-Edge$\ \quad$}}}   & Main Memory             & 1-Channel LP-DDR3-1600; Bandwidth = 25.6GB/s      \\ [1mm]
                                  & PEs                     & 64                                                \\ [1mm]
                                  & MAC Lanes               & 16 per PE                                         \\ [1mm]
                                  & Softmax Modules         & 4 per PE                                          \\ [1mm]
                                  & Batch Size              & 4                                                 \\ [1mm]
                                  & Buffer & Activation Buffer: 4MB; Weight Buffer: 8MB;       \\ [1mm]
                                  &                         & Mask Buffer: 1MB                                  \\ \midrule
\multirow{7}{*}{\rotatebox[origin=c]{90}{\textbf{AccelTran-Server$\quad$}}} & Main Memory             & 2-channel Mono. 3D RRAM; Bandwidth = 256GB/s \\ [1mm]
                                  & PEs                     & 512                                               \\ [1mm]
                                  & MAC Lanes               & 32 per PE                                         \\ [1mm]
                                  & Softmax Modules         & 32 per PE                                         \\ [1mm]
                                  & Batch Size              & 32 \\ [1mm] 
                                  & Buffer & Activation Buffer: 32MB; Weight Buffer: 64MB;     \\ [1mm]
                                  &                         & Mask Buffer: 8MB                                 \\ \bottomrule
\end{tabular}}
\label{tbl:acceltran_des_choices}
\end{table}

We now present various design choices for our proposed framework. We introduce two accelerators, namely 
AccelTran-Edge and AccelTran-Server. The first is for mobile/edge platforms with a limited energy budget. 
The second is aimed at cloud/server applications where throughput may be of utmost importance. Table~\ref{tbl:acceltran_des_choices} shows the associated 
design choices. We fixed the clock rate to 700 MHz based on 
the delay of all modules in the proposed architecture. We set the number of multipliers $M$ to 16. \textcolor{black}{We set IL = 4 and FL = 16.} As
mentioned in Section~\ref{sec:results_dataflow}, the dataflow \code{[b,i,j,k]} is the loop-unrolling scheme 
of choice. We set the tile sizes across \code{b}, \code{i}, and \code{j} to 1, 16, and 16, respectively. For the chosen RRAM process~\cite{rram_14nm} in AccelTran-Server, we implement the memory in two tiers above the main accelerator tier in order to fit it within the footprint area. However, 
different transformer models would generally have a unique set of hardware hyperparameters that are optimal 
for the given architecture. Thus, one can search for an optimal transformer-accelerator pair over a diverse set of transformer models~\cite{flexibert} and accelerator design choices~\cite{naas}.

\subsection{Evaluation Baselines}

We compare the performance of our proposed accelerator with many previously proposed baselines. For mobile 
platforms, we compare the inference of BERT-Tiny on AccelTran-Edge with off-the-shelf platforms that include 
Raspberry Pi 4 Model-B~\cite{rpi} that has the Broadcom BCM2711 ARM SoC, Intel Neural Compute Stick (NCS) 
v2~\cite{ncs} with its neural processing unit (NPU), and Apple M1 ARM SoC~\cite{apple_m1} with an 8-core CPU, an 8-core GPU, and 16 GB unified memory on an iPad (for easier evaluations, we performed experiments on a MacBook 
Pro laptop with the same SoC instead). For server-side platforms, we compare the inference of BERT-Base on 
AccelTran-Server with a modern NVIDIA A100 GPU (40GB of video RAM) and previously proposed accelerators, namely, 
OPTIMUS~\cite{optimus}, SpAtten~\cite{spatten}, and Energon~\cite{energon}. We chose the maximum batch size 
possible for each platform, based on its memory capacity. 

\textcolor{black}{To support inference on the Raspberry Pi, we implement the transformer models on an ARM distribution of the 
machine learning (ML) framework, PyTorch. We run transformer evaluation on the Intel NCS using the OpenVINO framework. Finally, 
for the Apple M1 
SoC, we use the Tensorflow-metal plug-in to exploit the CPU and its embedded GPU. We quantize all models to FP16 before 
running our experiments. We normalize the throughput, energy, and chip area to 14nm FinFET technology using scaling 
equations~\cite{technology_norm}. We use the inverter delays for different technology nodes as proxies for throughput 
normalization.}

\section{Experimental Results}
\label{sec:results}

In this section, we present the experimental results.

\subsection{Dynamic Inference with the Transformer}
\label{sec:results_dynatran}

We first present the results of our experiments for the DynaTran method.

\subsubsection{Comparing DynaTran with the Baseline}

\begin{figure}
    \centering
    \includegraphics[width=\linewidth]{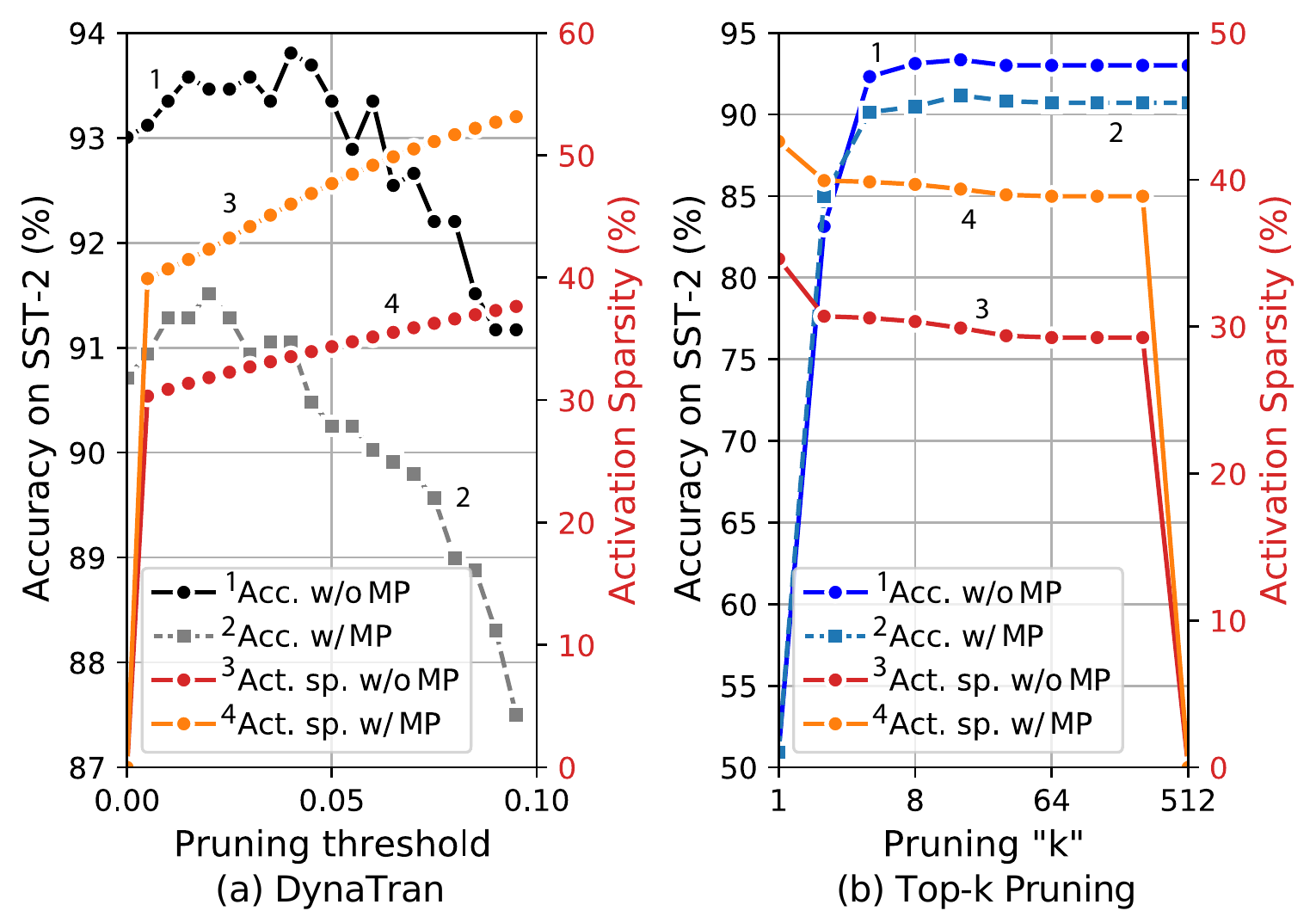}
    \caption{Accuracy on the SST-2 task and activation sparsity with (a) pruning threshold for DynaTran and 
(b) pruning ``$k$" for top-$k$ pruning.}
    \label{fig:acc_thresh}
\end{figure}

\begin{figure}
    \centering
    \includegraphics[width=0.95\linewidth]{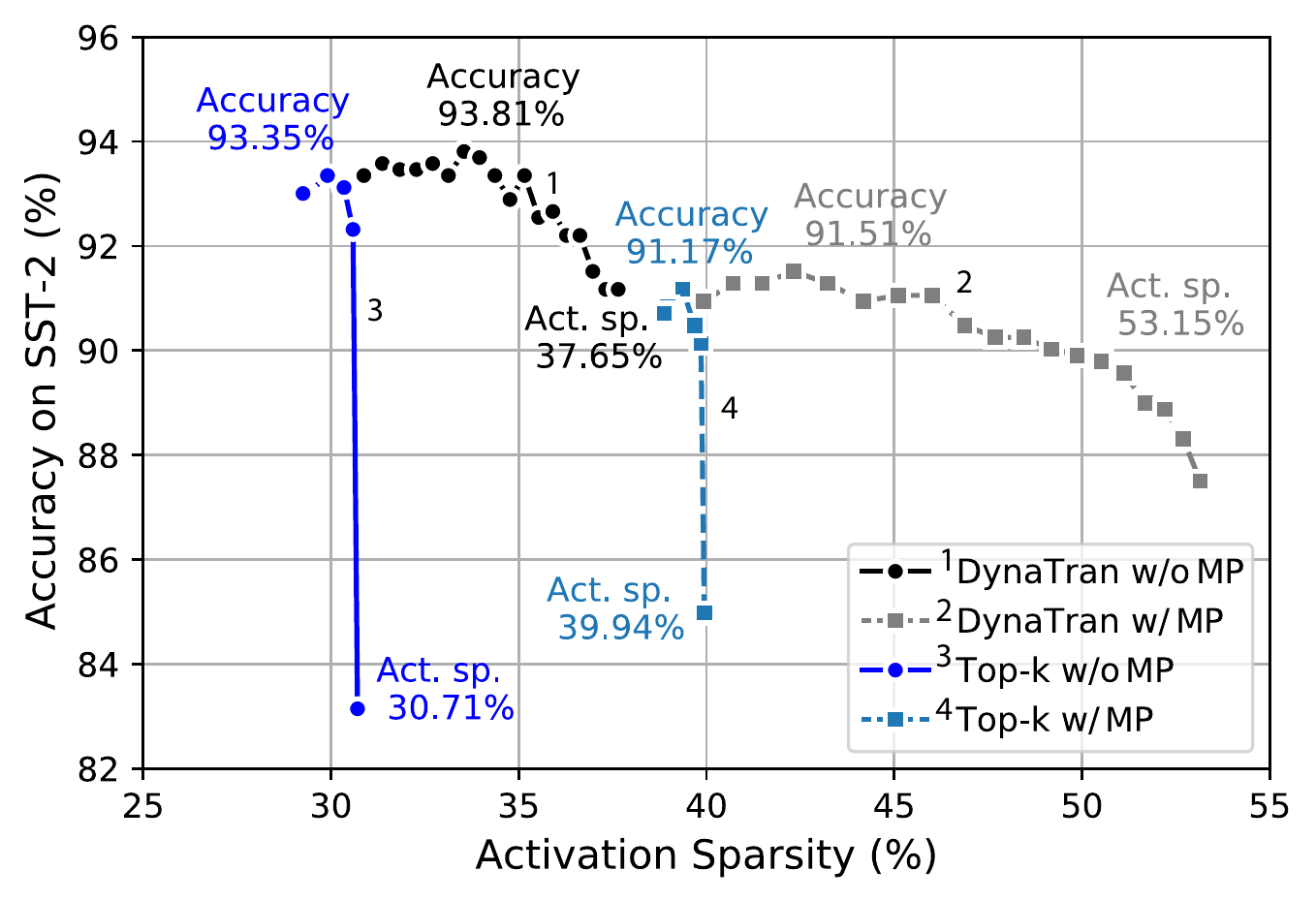}
    \caption{Accuracy on the SST-2 task with activation sparsity for DynaTran and top-$k$ methods. The 
annotations correspond to the maximum achieved accuracy or activation sparsity for each case.}
    \label{fig:acc_sp_dyna-vs-topk}
\end{figure}

Figs.~\ref{fig:acc_thresh} and \ref{fig:acc_sp_dyna-vs-topk} present the profiled accuracy curves for BERT-Base 
on the SST-2 task for DynaTran and top-$k$ pruning techniques. In Fig.~\ref{fig:acc_thresh}, we show the
effect of the pruning hyperparameters on sparsity. For DynaTran, the pruning threshold ($\tau$) is varied from 
0 to 0.1 and the activations are pruned based on the pruning threshold (see Section~\ref{sec:dynatran}). 
For top-$k$ pruning, we change $k$ in powers of two in order to see the effect of \emph{net} activation sparsity, 
i.e., the sparsity in all activations rather than only the attention scores. Further, we also test pre-pruned 
models to see the impact on net activation sparsity when weights are also pruned. For this, we use the BERT-Base 
model pruned using the MP algorithm~\cite{movement_pruning}. Using MP results in a higher activation sparsity 
(since the activations formed by matrix multiplications with weights are sparser when the weights are also sparse), 
but at the cost of lower accuracy. As also observed in previous works~\cite{energon}, both DynaTran and 
top-$k$ methods see an initial increase in accuracy before a drop, as the sparsity increases. This could be 
attributed to the over-parameterization of the BERT model~\cite{bert_overparameterized} and the corresponding pruning 
method acting as a regularizer, thus giving a slightly higher validation performance.

We see similar results for other models and datasets. We store geometric mean curves, like the ones presented here, 
in the internal register of the DynaTran module with a low memory footprint. For the required activation 
sparsity, or even accuracy, we obtain the corresponding pruning threshold through the threshold calculator in the 
DynaTran module (explained in Section~\ref{sec:dynatran_module}) to implement the desired dynamic inference.

Fig.~\ref{fig:acc_sp_dyna-vs-topk} plots accuracy curves against activation sparsity for the DynaTran and top-$k$ 
methods with and without MP. We obtain these curves from those in Fig.~\ref{fig:acc_thresh} by plotting 
accuracy against the corresponding resultant activation sparsity for every pruning threshold ($\tau$) or the 
pruning $k$, as per the chosen method. We can see the trend of a slight increase in accuracy here as well. DynaTran 
achieves a higher accuracy (0.46\% higher for BERT-Base without MP and 0.34\% higher with MP) and a higher possible activation sparsity without much accuracy loss for both cases, i.e., with and without MP. 
For the same accuracy (the highest achievable by top-$k$), DynaTran enables 1.17$\times$ and 1.20$\times$ higher 
activation sparsity for each case, respectively. On the other hand, DynaTran can achieve up to 1.33$\times$ 
(1.23$\times$) higher sparsity in absolute terms without MP (with MP). Here, we use $\tau < 0.1$, which yields
reasonable accuracy values.

\begin{figure}
    \centering
    \includegraphics[width=0.95\linewidth]{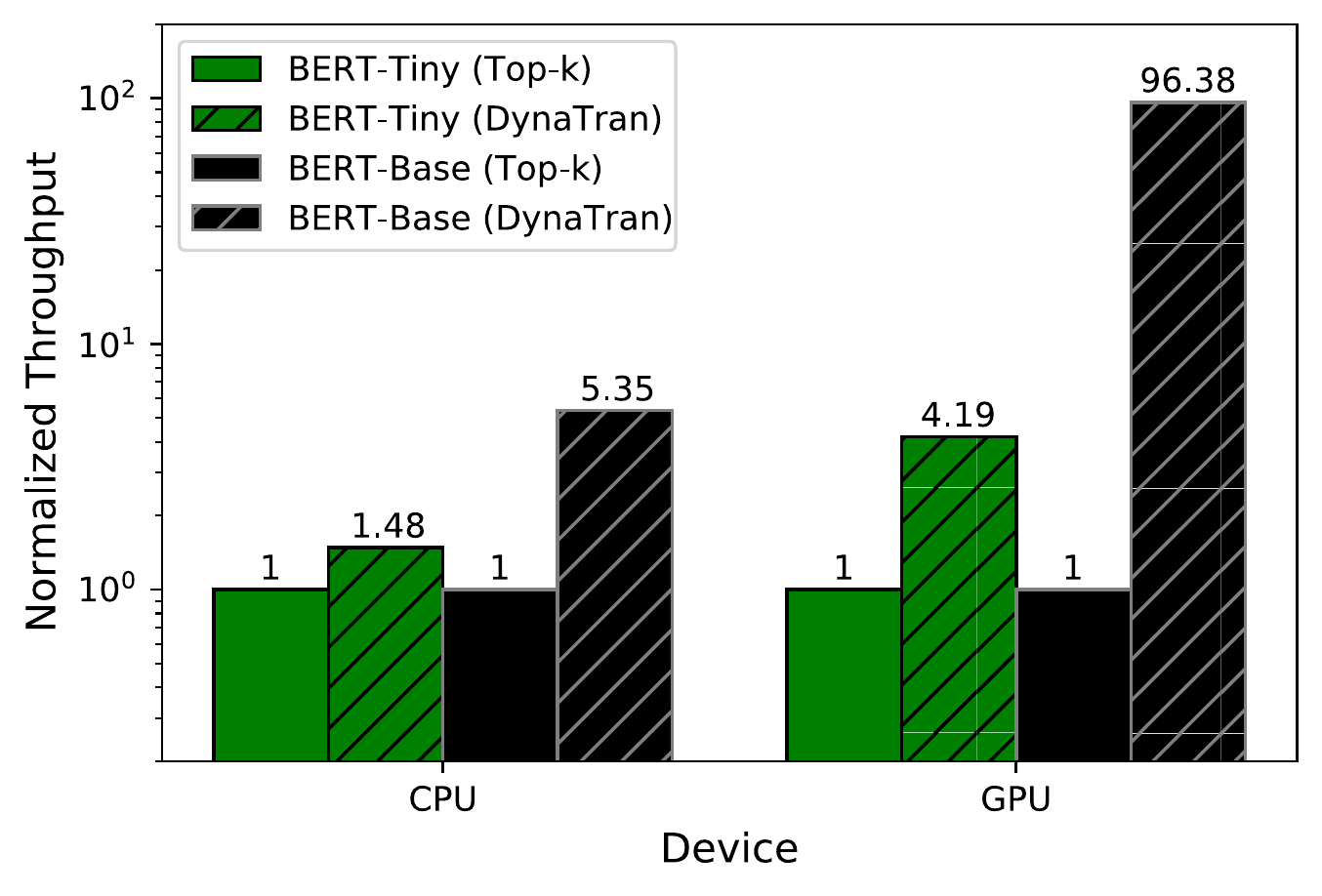}
    \caption{Normalized throughput of DynaTran compared with the top-$k$ method on a CPU and a GPU. Annotations are presented over each bar.}
    \label{fig:throughput_dyna-vs-topk}
\end{figure}

We now compare the compute cost of the top-$k$ method with that of DynaTran. Fig.~\ref{fig:throughput_dyna-vs-topk} 
shows the normalized throughputs of the two methods for BERT-Tiny and BERT-Mini on two devices. These are a 
2.6 GHz AMD EPYC Rome CPU with 128 cores and 768GB memory and an A100 GPU with 40GB VRAM. DynaTran achieves up to 
96.38$\times$ higher throughput on the GPU and up to 5.35$\times$ higher throughput on the CPU. This 
is due to the use of low-overhead comparators with a pre-determined threshold. Even with the specialized top-$k$ engine used in SpAtten 
and the approximation scheme used in Energon~\cite{energon}, they use more than one clock cycle, 
whereas DynaTran uses just one clock cycle. This is because the threshold calculator only needs a simple look-up operation and the comparators can execute within a clock cycle.

\subsubsection{Testing if Weight Pruning is Effective in DynaTran}
\label{sec:results_dynatran_wp}

\begin{figure}
    \centering
    \includegraphics[width=\linewidth]{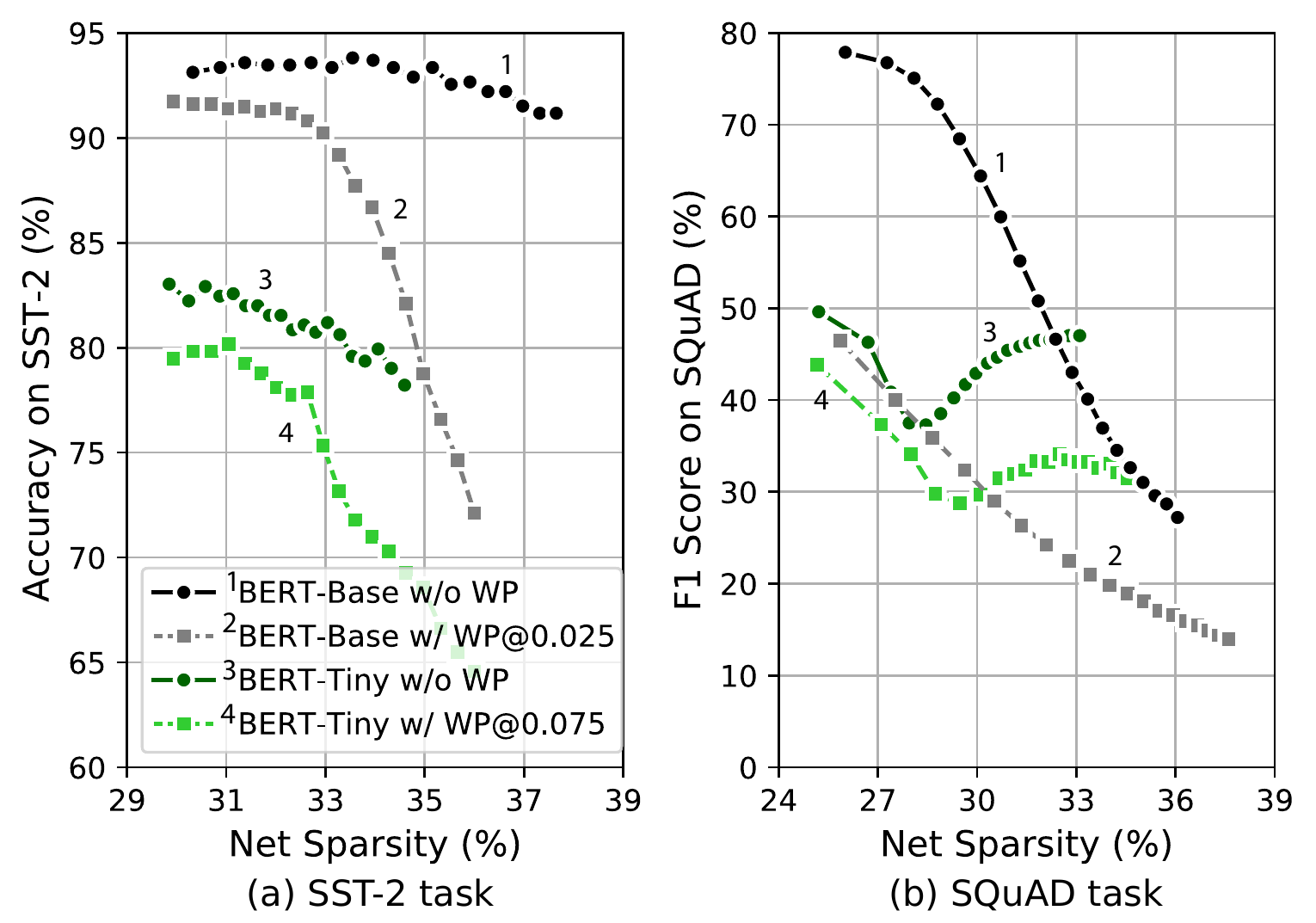}
    \caption{Accuracy/F1-score plotted against net sparsity on the (a) SST-2 and (b) SQuAD benchmarks. In
DynaTran, WP was implemented with a fixed threshold.}
    \label{fig:weight_pruning}
\end{figure}

DynaTran implements magnitude-based pruning of all activations at runtime. However, we can also leverage it to prune 
model weights before running the transformer. We call this weight pruning (WP) since we only prune the transformer 
weights. In this approach, we do not need downstream training, as opposed to MP, which iteratively trains model weights 
while also pruning them. Fig.~\ref{fig:weight_pruning} presents the accuracies and F1-scores on the SST-2 and SQuAD 
datasets, respectively, with and without the use of WP. Net sparsity represents the combined sparsity of weights 
and activations. WP results in slightly higher net sparsity, however, with a significant loss in performance. The 
high ratio of activations compared to weights (see Fig.~\ref{fig:mem_req}) results in only marginal gains in net 
sparsity. Hence, we do not employ WP in DynaTran. We use movement-pruned models instead, resulting in high weight 
and activation sparsities (with DynaTran) at negligible performance loss.

\subsection{Dataflows and Data Reuse}
\label{sec:results_dataflow}

\begin{figure}
    \centering
    \includegraphics[width=\linewidth]{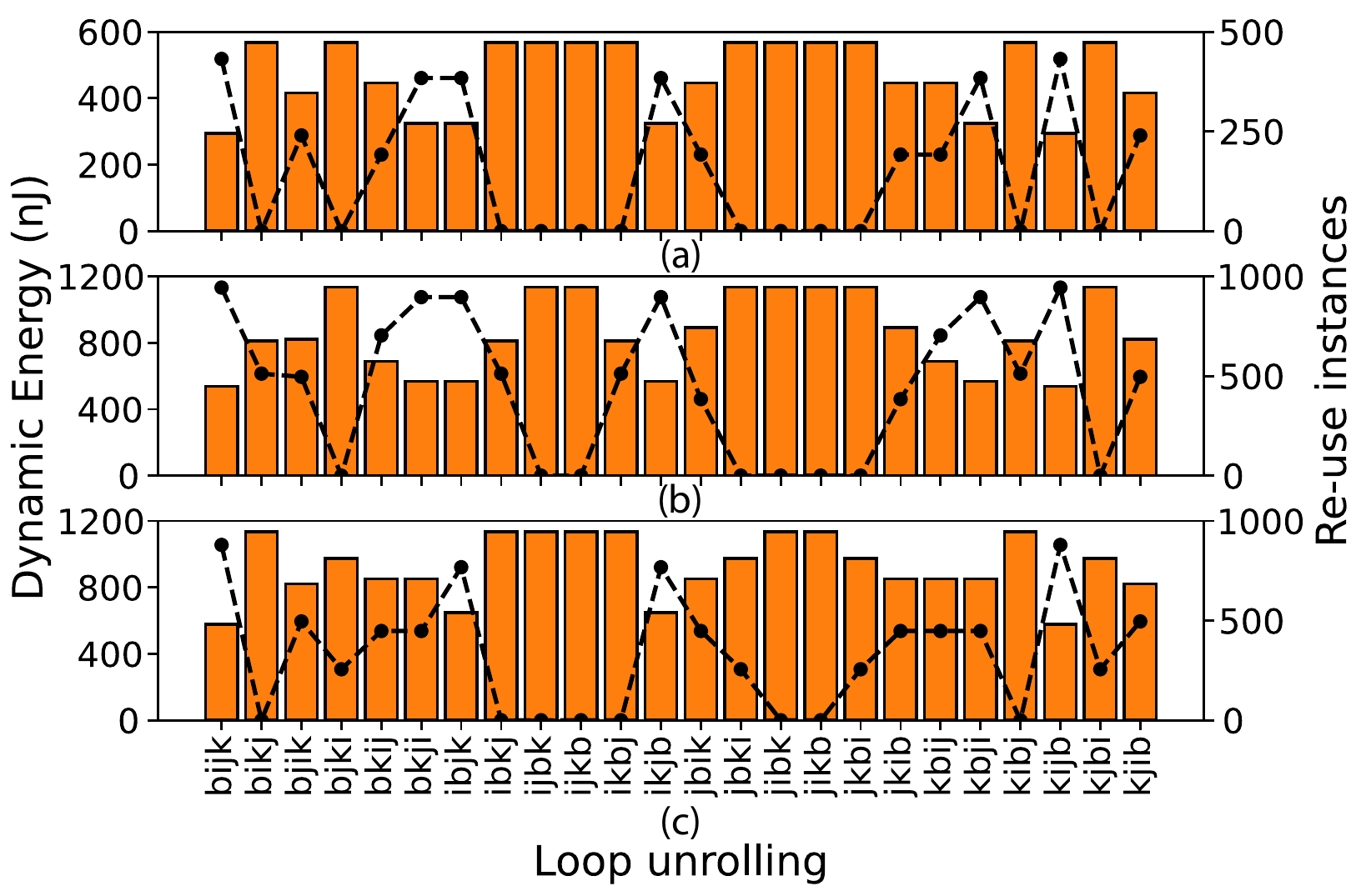}
    \caption{Comparison of energy and reuse instances for all 24 dataflows under three matrix multiplication 
($\mathbf{W} \times \mathbf{A}$) scenarios: (a) $\mathbf{W} \in \mathbb{R}^{4 \times 64 \times 64}, 
\mathbf{A} \in \mathbb{R}^{4 \times 64 \times 64}$, (b) $\mathbf{W} \in \mathbb{R}^{4 \times 64 \times 64}, \mathbf{A} 
\in \mathbb{R}^{4 \times 64 \times 128}$, and (c) $\mathbf{W} \in \mathbb{R}^{4 \times 128 \times 64}, \mathbf{A} 
\in \mathbb{R}^{4 \times 64 \times 64}$. Bar plots represent dynamic energy and dashed lines represent reuse instances.}
    \label{fig:dataflow}
\end{figure}

We can pass on different tiles to available resources based on the four for-loops shown in 
Fig.~\ref{fig:tiling_dataflow}. We can arrange these four for-loops in $^4P_4 = $ 24 ways without changing the 
output. However, based on the compute resource constraints, different loop-unrolling strategies, or dataflows, can 
result in the reuse of local tiled weights or activations. Fig.~\ref{fig:dataflow} compares these dataflows for 
various matrix multiplication operations. The multiplication, $\mathbf{W} \times \mathbf{A}$, is carried out
using four MAC lanes in this simple example. We observe that dynamic energy is minimized by dataflows 
\code{[b,i,j,k]} and \code{[k,i,j,b]}. We use the former dataflow for subsequent experiments. These two dataflows also 
have maximum reuse instances for all three matrix multiplications. A reuse instance indicates if a weight or activation 
tile is reused in the internal register of a MAC lane. Many dataflows have the same energy and reuse instances due to 
symmetry. Since AccelTran hides data transfer overheads, due to the optimized control 
flow, the net latency is the same for all dataflows (this also results in the same leakage energy).

Next, we test the effect of the different dataflows on real-world traces with the BERT-Tiny model on AccelTran-Edge. 
However, we observed negligible energy differences among the dataflows. This could be attributed to massive parallelization being at odds with data reuse. For instance, to reuse the same set of tiled weights in a 
PE's register, the next operation using those weights would have to be assigned to the same PE rather than exploit 
other free PEs, thus limiting parallelization. Hence, as per Fig.~\ref{fig:dataflow}, the advantages of data reuse can only be exploited in highly resource-constrained accelerators. 

\subsection{Design Space Exploration}

\begin{figure}
    \centering
    \includegraphics[width=0.9\linewidth]{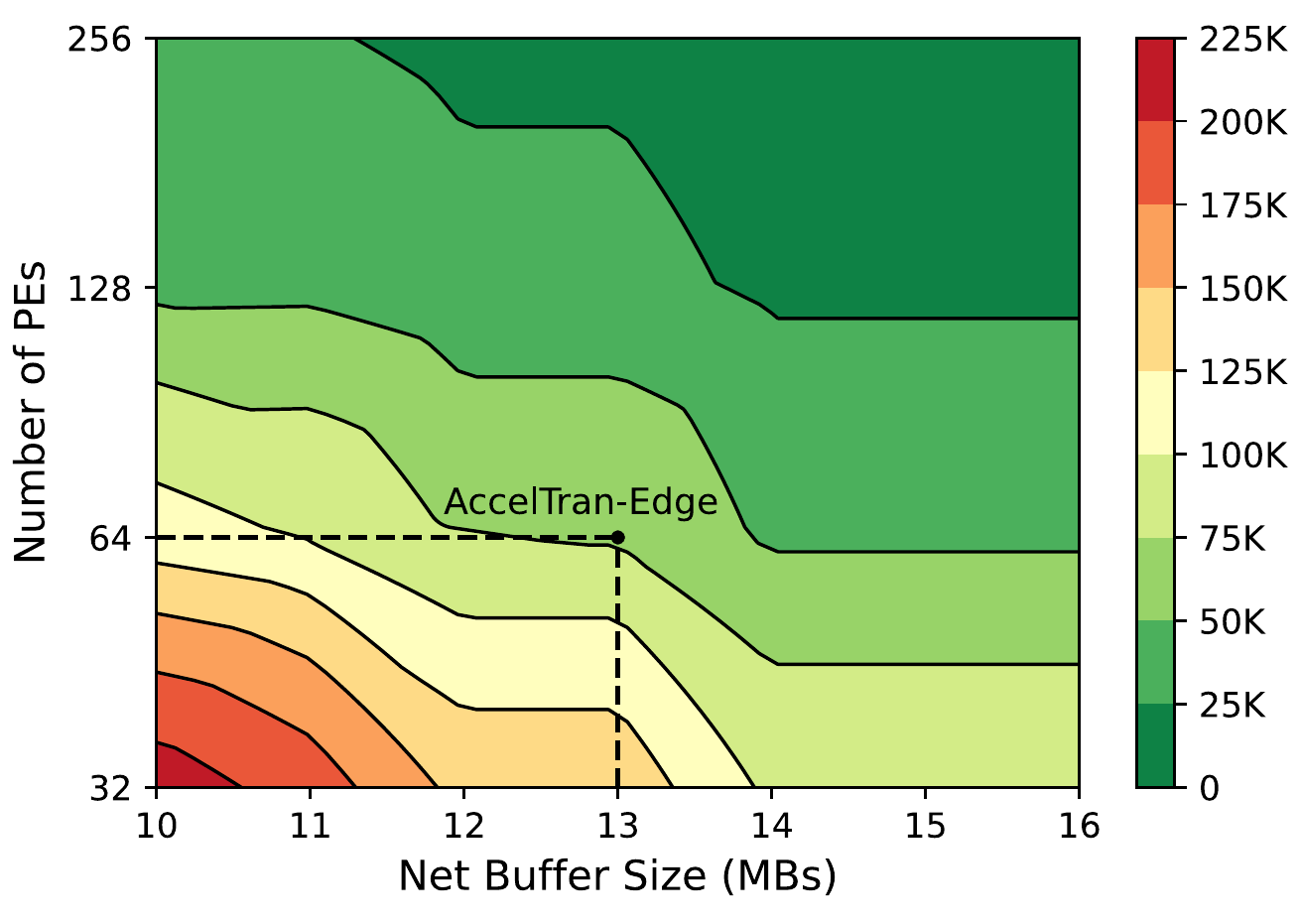}
    \caption{Number of stalls with hardware resources.}
    \label{fig:exploration}
\end{figure}

Fig.~\ref{fig:exploration} shows a plot of the number of compute and memory stalls when evaluating BERT-Tiny with 
different number of PEs and buffer sizes. We use a 4:8:1 size ratio for the activation, weight, and mask buffers. We found this ratio to be close to the optimal based on empirical studies on memory access patterns for the BERT-Tiny model. 
Next, we sweep the net buffer size from 10MB to 16MB. Finally, we choose the following number of PEs: 32, 64, 128, and 256. 
The figure shows that the number of compute stalls gradually increases as both the number of PEs and buffer size are 
reduced. We justify this as follows. 

\begin{table*}[]
\caption{Area, theoretical peak TOP/s, and minimum main memory requirements, along with power consumption
breakdown for different parts of the proposed accelerator architectures. The LP mode for AccelTran-Edge is 
also considered.}
\centering
\resizebox{0.9\linewidth}{!}{
\begin{tabular}{@{\hskip 0.2in}l@{\hskip 0.2in}|@{\hskip 0.2in}c@{\hskip 0.2in}c@{\hskip 0.2in}c@{\hskip 0.2in}|@{\hskip 0.2in}c@{\hskip 0.2in}c@{\hskip 0.2in}c@{\hskip 0.2in}c@{\hskip 0.2in}}
\toprule
                         &            &        &                 & \multicolumn{4}{c}{\textbf{Power Breakdown (W)}} \\
\textbf{Accelerator/Operation}    & \textbf{Area (mm$^2$)} & \textbf{TOP/s} & \textbf{Main Mem. (MB)} & \textbf{PEs}    & \textbf{Buffers}  & \textbf{Main Mem.}  & \textbf{Total}  \\ \midrule
AccelTran-Server         & 1950.95    & 372.74 & 3467.30         & 48.25  & 10.40    & 36.86      & 95.51  \\ [1mm]
AccelTran-Edge           & 55.12      & 15.05  & 52.88           & 3.79   & 0.08     & 2.91       & 6.78   \\ [1mm]
AccelTran-Edge (LP mode) & 55.12      & 7.52   & 52.88           & 2.31   & 0.05     & 1.77       & 4.13   \\ \bottomrule
\end{tabular}}
\label{tbl:hw_perf_details}
\end{table*}

A lower number of PEs results in increased compute stalls since the compute operations have to wait for resources to 
free up in order to execute them, limiting available parallelization. In addition, a small buffer size results in memory stalls 
since memory store operations have to wait for the corresponding compute operations to finish before the current 
activations or weights, initially required by those compute operations, can be evicted from the buffer. 
Fig.~\ref{fig:exploration} shows the chosen point for AcelTran-Edge. This set of design choices (64 PEs and 13MB net 
buffer size) represents a reasonable trade-off between the number of stalls (that directly increase latency) and 
hardware resources (that directly increase area and power consumption). An automatic hardware-software co-design approach~\cite{codebench} could also \emph{efficiently} test different buffer sizes, along with the corresponding 
ratios that may be 
optimal for each transformer model. We defer this automated co-design method to future work.

\subsection{Hardware Performance and Utilization}

\begin{figure}
    \centering
    \includegraphics[width=\linewidth]{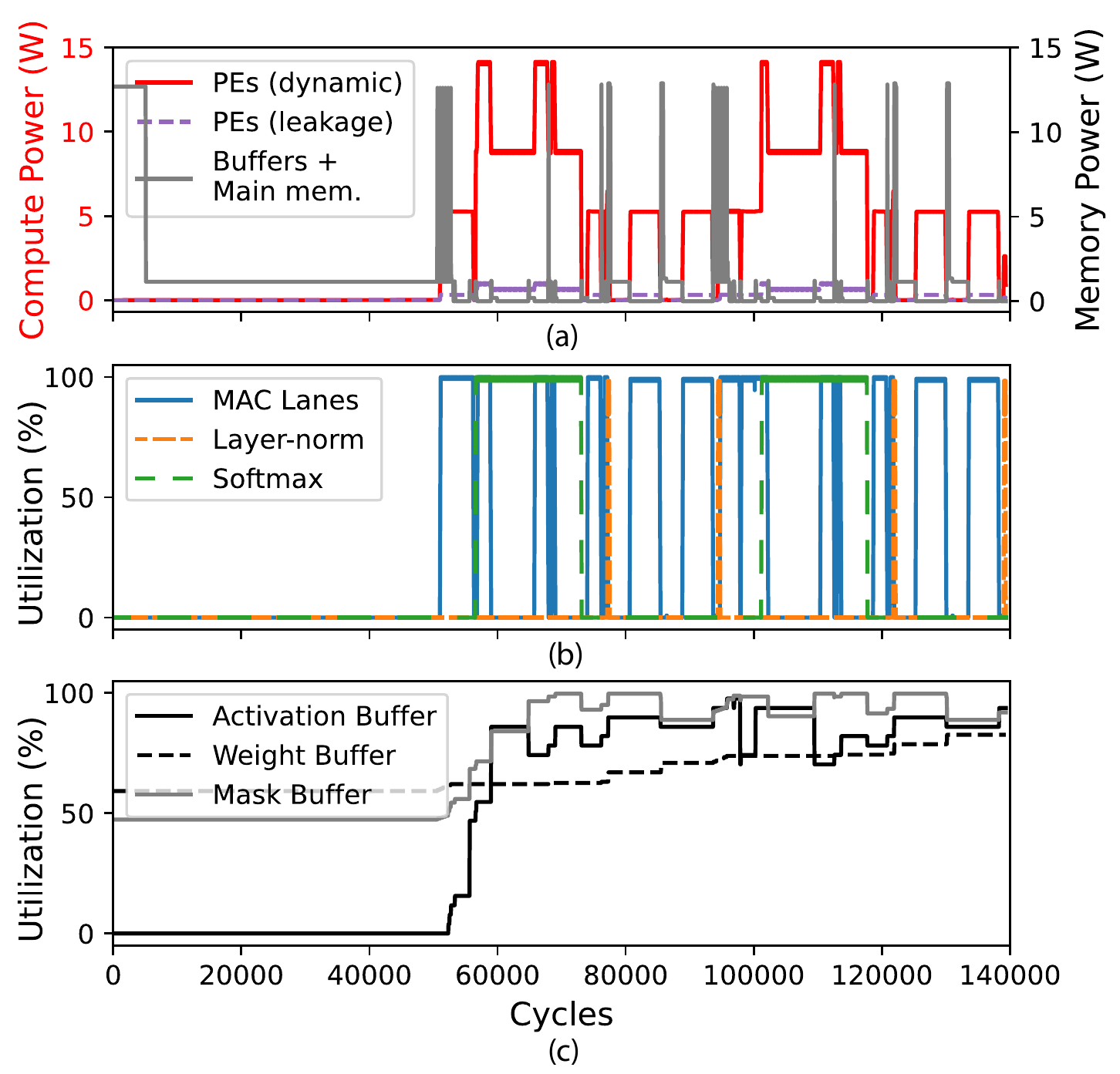}
    \caption{Evaluation of BERT-Tiny on AccelTran-Edge: (a) power consumption, (b) resource utilization of compute modules, and (c) resource utilization of buffers.}
    \label{fig:bert_tiny_edge_metrics}
\end{figure}

Fig.~\ref{fig:bert_tiny_edge_metrics} shows the power consumption and resource utilization of BERT-Tiny on
AccelTran-Edge during inference of one batch. Hardware utilization remains at zero until around 51K cycles (see Fig.~\ref{fig:bert_tiny_edge_metrics}(b)) when the 
accelerator loads the word and position embeddings into the weight buffer (accounting for around 60\% of the weight 
buffer). However, these load operations only occur once and subsequent transformer evaluations on different sequences reuse 
these embeddings. The rest of the process sees high utilization of MAC lanes or softmax modules. At certain times, the accelerator uses
both MAC lanes and softmax modules due to the staggered implementation of attention head operations. The leakage power is low, as we show in Fig.~\ref{fig:bert_tiny_edge_metrics}(a), due to the power-gating of unused modules. Buffer usage drops suddenly, in Fig.~\ref{fig:bert_tiny_edge_metrics}(c), at certain instances when data are evicted in order to make space for new data for the active compute operations.

Table~\ref{tbl:hw_perf_details} shows the hardware performance measures for the proposed accelerator architectures, 
namely AccelTran-Server and AccelTran-Edge, along with a low-power (LP) mode that we support for AccelTran-Edge. 
The LP mode only works with half of the compute hardware at any given time, resulting in lower net power draw, which 
is often a constraint in edge devices that rely on a battery source. We show the chip area first. AccelTran-Server 
is a massive chip with an area of 1950.95 mm$^2$, although still lower than that of the A100 GPU (3304 mm$^2$ normalized to a 14nm 
process~\cite{technology_norm}). This can reduce the yield. However, we can leverage intelligent placement of PEs and binning to improve 
apparent yield rates~\cite{binning}. We also show the tera-operations per second (TOP/s) performance measure for both architectures. AccelTran-Server can theoretically achieve a peak performance of 372.74 TOP/s, 
assuming all compute modules are operational simultaneously. We also present the minimum main memory size required 
for each accelerator. The net size of the embeddings and weights for BERT-Base and BERT-Tiny are 3467.30MB and 
52.88MB (assuming a conservative 50\% weight sparsity ratio~\cite{movement_pruning}), respectively. However, transformer 
evaluation does not require all 
weights at any given time. Thus, the weight buffer can be much smaller. Similarly, even though 
the net size of activations is much higher (see Fig.~\ref{fig:mem_req}), we can use a much smaller activation buffer. Finally, we present the power breakdowns for both the accelerators and the LP mode for AccelTran-Edge. The LP 
mode reduces power consumption by 39.1\%, while lowering throughput by 38.7\%, for BERT-Tiny.

\begin{figure}
    \centering
    \includegraphics[width=\linewidth]{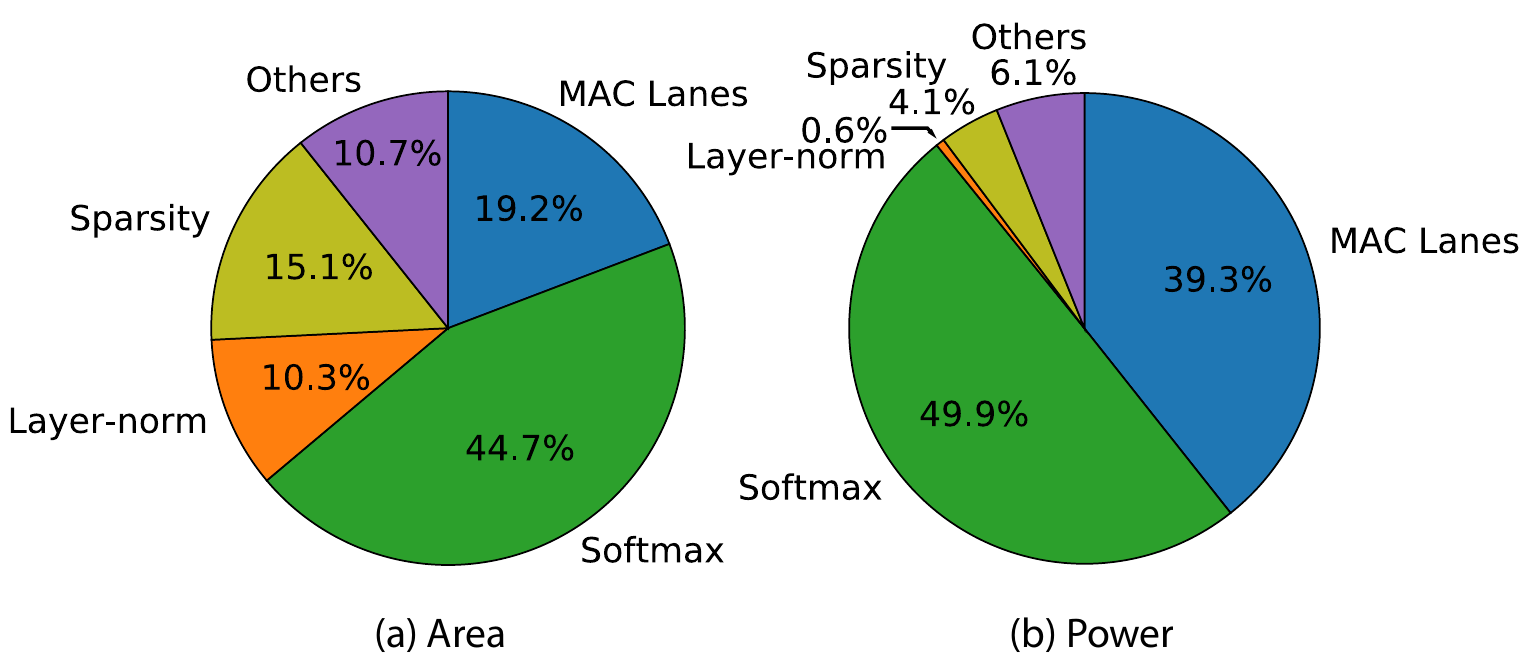}
    \caption{Breakdown of (a) area and (b) power consumption by compute modules in AccelTran-Edge.}
    \label{fig:area_power_bkdwn}
\end{figure}

Fig.~\ref{fig:area_power_bkdwn} shows the area and power breakdowns for different compute modules in AccelTran-Edge. 
The 1024 MAC lanes only take up 19.2\% of the area, while the specialized 256 softmax and 64 layer-norm modules take 
up 44.7\% and 10.3\% of the area, respectively. Pre- and post-compute sparsity modules comprise 15.1\% area, while 
the dataflow, the DynaTran modules, and the DMA occupy 10.7\% of the chip area. Fig.~\ref{fig:area_power_bkdwn}(b) shows 
the average power breakdown. Since most operations in the transformer involve matrix multiplication or softmax, 
they also draw most of the power (39.3\% for MAC lanes and 49.9\% for softmax modules). The high power consumption of the 
softmax modules can be attributed to the calculation of the exponential sum over the entire tile in a parallel manner.

\subsection{Effect of Sparsity on Throughput and Energy}

\begin{figure}[t]
    \centering
    \includegraphics[width=\linewidth]{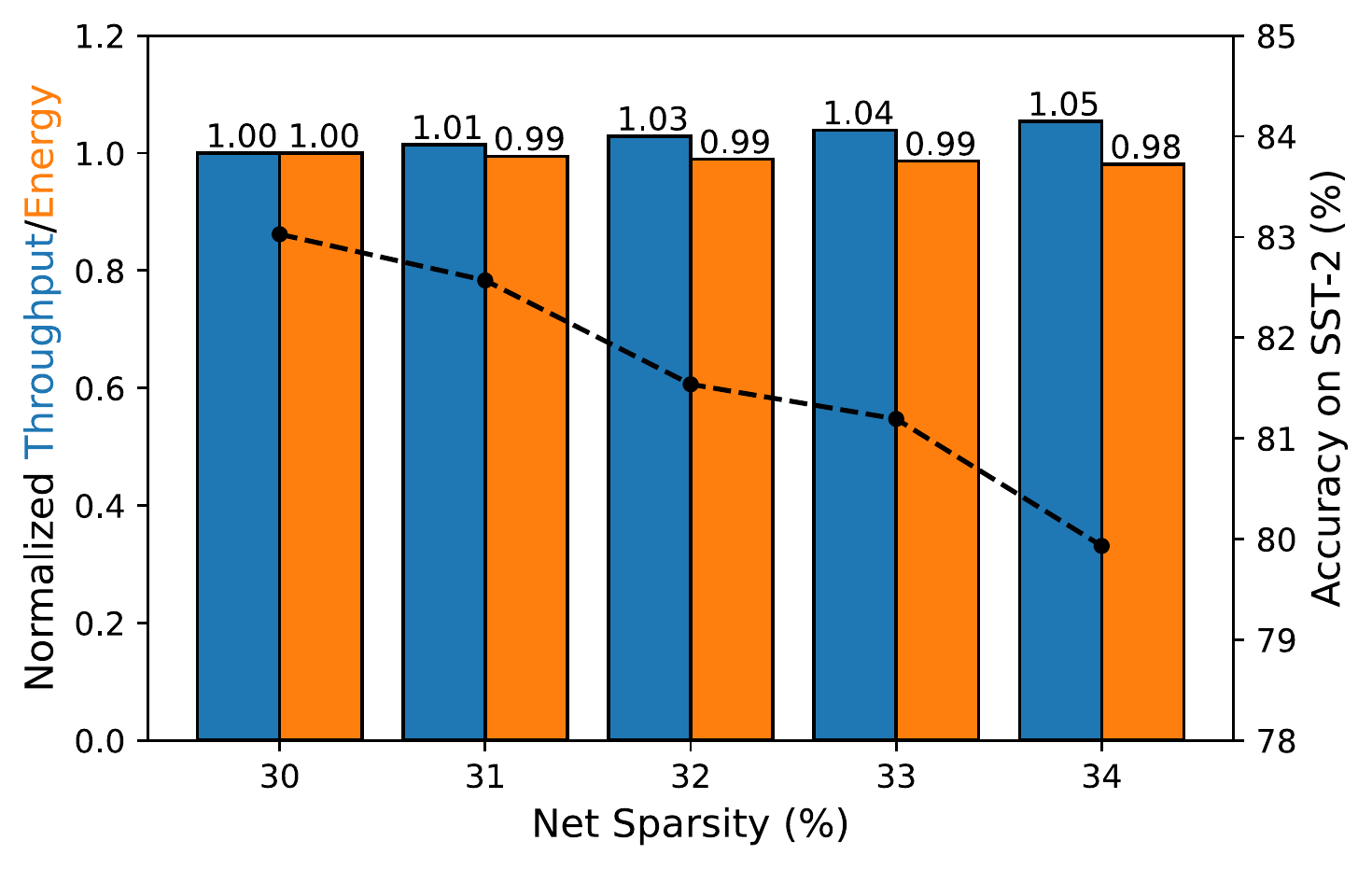}
    \caption{Effect of sparsity on throughput and energy consumption. BERT-Tiny is simulated on AccelTran-Edge. 
Normalized throughput and energy are shown as bar plots on the left, and accuracy is shown as a dashed line plot on 
the right.}
    \label{fig:sparsity}
\end{figure}

Fig.~\ref{fig:sparsity} shows the effect of increasing sparsity on accelerator throughput and energy consumption. 
As the net sparsity increases from 30\% to 34\% for the BERT-Tiny model (with a conservative 50\% weight sparsity 
estimate and accordingly tuned DynaTran's thresholds), throughput improves by 5\% whereas energy consumption drops 
by 2\%, when implemented on AccelTran-Edge. Here, accuracy drops by only 3\% due to the low performance 
loss of DynaTran.

\subsection{Performance Improvements}

\begin{figure}[!t]
    \centering
    \includegraphics[width=\linewidth]{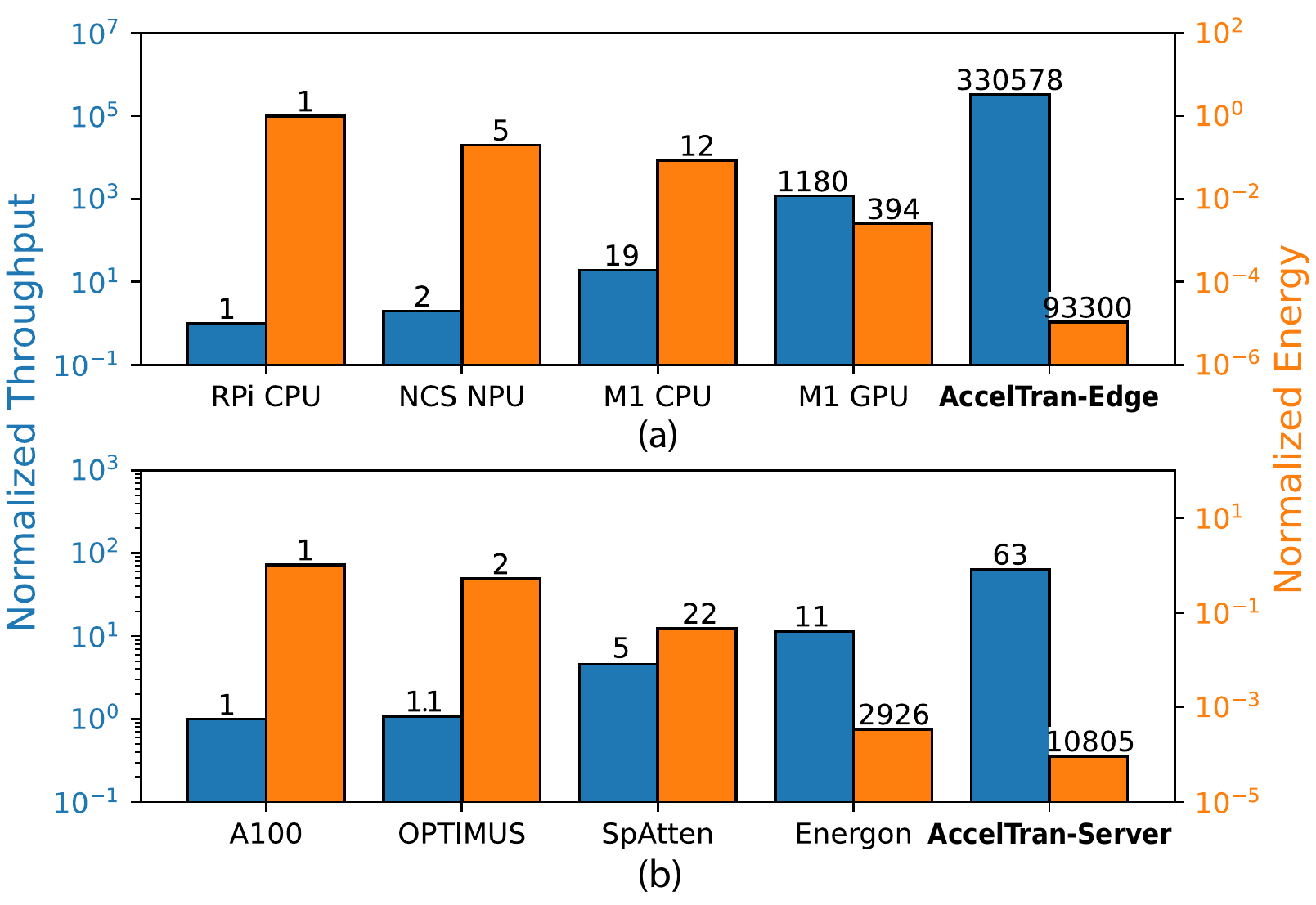}
    \caption{Normalized throughput (left) and energy (right) comparison for AccelTran with baseline platforms targeted 
at (a) edge and (b) server applications.}
    \label{fig:baselines}
\end{figure}

Fig.~\ref{fig:baselines} shows performance comparisons of AccelTran architectures with baseline platforms. For edge 
applications, we compare the inference of BERT-Tiny on AccelTran-Edge with that on Raspberry Pi CPU, Intel NCS NPU, 
M1 CPU, and M1 GPU. AccelTran-Edge achieves 330,578$\times$ higher throughput at 93,300$\times$ lower energy 
consumption relative to Raspberry Pi. On the server side, we compare the performance of BERT-Base on AccelTran-Server 
with that of A100 GPU and some recently proposed accelerators, namely, OPTIMUS~\cite{optimus}, SpAtten~\cite{spatten}, 
and Energon-Server~\cite{energon}. The throughput and energy values for SpAtten and Energon are 
normalized with respect to the A100 GPU. AccelTran-Server achieves 63$\times$ (5.73$\times$) higher throughput at 
10,805$\times$ (3.69$\times$) lower energy consumption when compared to off-the-shelf A100 GPU (state-of-the-art 
Energon co-processor). \textcolor{black}{These gains can be attributed to the execution of the DynaTran algorithm at runtime along with sparsity-aware modules that skip ineffectual computations. The specialized softmax and layer-norm modules also speed up the respective operations, otherwise implemented as matrix multiplications in the A100. Further, monolithic-3D RRAM has much lower data-retrieval latency than HBM in the A100. These contributions enable AccelTran to achieve high throughput gains over the A100 GPU. We study the effects of these contributions next.}

\begin{table}[t]
\caption{Ablation analysis for inference of BERT-Tiny on AccelTran-Server}
\centering
\begin{tabular}{@{}l|ccc@{}}
\toprule
\textbf{Accelerator Configuration}         & \begin{tabular}[c]{@{}c@{}}\textbf{Throughput} \\ \textbf{(seq/s)}\end{tabular} & \begin{tabular}[c]{@{}c@{}}\textbf{Energy}\\ \textbf{(mJ/seq)}\end{tabular} & \begin{tabular}[c]{@{}c@{}}\textbf{Net Power} \\ \textbf{(W)}\end{tabular} \\ \midrule
\textbf{AccelTran-Server} & \textbf{172,180} & \textbf{0.1396} & 24.04 \\ [1mm]
$\quad$w/o DynaTran & 93,333 & 0.1503 & \textbf{14.03} \\ [1mm]
$\quad$w/o MP & 163,484 & 0.2009 & 32.85 \\ [1mm]
$\quad$w/o Sparsity-aware modules & 90,410 & 0.2701 & 24.43 \\ [1mm]
$\quad$w/o Monolithic-3D RRAM & 88,736 & 0.1737 & 15.42 \\ \bottomrule
\end{tabular}
\label{tbl:acceltran_server_ablation}
\end{table}

\subsection{Ablation Analysis}

\begin{table*}[]
\caption{Comparison of our proposed AccelTran framework with related works along different dimensions. 
$^*$Energon is not an accelerator but a co-processor.}
\centering
\begin{tabular}{@{\hskip 0.2in}l@{\hskip 0.2in}|@{\hskip 0.2in}c@{\hskip 0.2in}c@{\hskip 0.2in}c@{\hskip 0.2in}c@{\hskip 0.2in}c@{\hskip 0.2in}c@{\hskip 0.2in}c@{\hskip 0.2in}}
\toprule
\textbf{Work} & \begin{tabular}[c]{@{}c@{}}\textbf{Transformer} \\ \textbf{Acceleration}\end{tabular} & \begin{tabular}[c]{@{}c@{}}\textbf{ASIC-based} \\ \textbf{Acceleration}\end{tabular} & \begin{tabular}[c]{@{}c@{}}\textbf{Monolithic} \\ \textbf{3D-RRAM}\end{tabular} & \begin{tabular}[c]{@{}c@{}}\textbf{Tiled} \\ \textbf{Mat. Mult.}\end{tabular} & \begin{tabular}[c]{@{}c@{}}\textbf{Dataflow} \\ \textbf{Support}\end{tabular} & \textbf{Sparsity-aware} & \begin{tabular}[c]{@{}c@{}}\textbf{Dynamic} \\ \textbf{Inference}\end{tabular} \\ \midrule
SPRING~\cite{spring} &  & \cmark & \cmark &  &  & \cmark & \\ \hline
FTRANS~\cite{ftrans} & \cmark &  &  &  &  &  & \\\hline
FPGA Transformer~\cite{fpga_accelerator_1} & \cmark &  &  &  &  &  &  \\ \hline
A$^3$~\cite{a3} & \cmark & \cmark &  &  &  &  & \cmark \\ \hline
iMTransformer~\cite{imtransformer} & \cmark & \cmark &  &  &  & \cmark &  \\ \hline
OPTIMUS~\cite{optimus} & \cmark & \cmark &  &  &  & \cmark & \\ \hline
SpAtten~\cite{spatten} & \cmark & \cmark &  &  &  & \cmark & \cmark \\ \hline
Energon$^*$~\cite{energon} & \cmark &  &  &  &  & \cmark & \cmark \\ \hline
\textbf{AccelTran (Ours)} & \cmark & \cmark & \cmark & \cmark & \cmark & \cmark & \cmark \\ \bottomrule
\end{tabular}
\label{tbl:baseline_comparison}
\end{table*}

Table~\ref{tbl:acceltran_server_ablation} presents an ablation analysis for the inference of BERT-Tiny on 
AccelTran-Server. The first row corresponds to the selected AccelTran configuration as per 
Table~\ref{tbl:acceltran_des_choices}, with 50\% weight sparsity implemented through MP and 50\% activation sparsity 
at runtime through DynaTran. The second row corresponds to the case not leveraging DynaTran. Then, we test the 
accelerator when the BERT model is not weight-pruned using MP. Third, we test it without employing the pre- and post-sparsity 
modules to skip ineffectual MAC operations. Finally, we present results when AccelTran-Server 
utilizes an off-chip LP-DDR3 DRAM instead of a high bandwidth monolithic-3D RRAM. Although the use of DRAM leads to a lower net 
average power consumption than when monolithic-3D RRAM is used, its total energy is higher due to a much lower 
throughput. 

\section{Discussion}
\label{sec:discussion}

In this section, we discuss the implications of the proposed accelerator in the field of machine learning (ML) 
acceleration and future work directions.

\subsection{\textcolor{black}{Dynamic Inference with Transformers}}

\textcolor{black}{Previous works leverage complex pruning mechanisms, like top-$k$ pruning, MP, etc. Implementing such pruning steps at runtime significantly slows down transformer evaluation. This has been a bottleneck in the widespread adoption of transformers on mobile platforms. In this work, we proposed a lightweight but powerful pruning mechanism: DynaTran. In essence, DynaTran implements magnitude-based pruning. However, we propose many novelties beyond vanilla magnitude-based pruning in terms of the algorithm and specialized hardware in order to obtain high gains relative to previous works. First, unlike previous works~\cite{kwon2022a, behnke-heafield-2020-losing}, we prune not only the weights but also all the activations, which are formed at runtime. Second, we store pre-profiled curves in the internal register of the DynaTran module. The threshold calculator selects the threshold for pruning at runtime based on user-defined constraints on accuracy or throughput. This enables dynamic adjustment of the desired accuracy or throughput at runtime (see trade-off shown in Fig.~\ref{fig:sparsity}). Third, the specialized DynaTran hardware module implements the algorithm in a single clock cycle, enabling high gains in throughput and reducing the bottlenecking effects of model pruning. Finally, DynaTran can easily incorporate any pre-processed weight pruning strategy~\cite{kwon2022a, behnke-heafield-2020-losing} into its pipeline. In our work, we show how we leverage movement-pruned models to enable higher sparsity in weights and activations. DynaTran results in better accuracy than the top-$k$ hardware-aware pruning mechanism and significantly improves throughput.}

\subsection{ML Accelerators}

Various proposed ML accelerators target specific architectures. CNN accelerators~\cite{eyeriss, spring, dadiannao, 
cnvlutin} focus on the convolution operation. Some works exploit sparsity in CNN models to reduce computation and 
memory footprint~\cite{spring, cambricon-x, cambricon-s}. Certain works also exploit dynamism in model representation 
to minimize performance loss while leveraging low-bit computation. Two recent works, DUET~\cite{duet} and 
Energon~\cite{energon}, employ dynamic mixed-precision computation. On the other hand, SPRING~\cite{spring} implements 
stochastic rounding~\cite{stochastic_rounding} with a fixed-precision format to maintain accuracy during training of CNNs. These 
extensions are orthogonal to the AccelTran framework and can easily be added to boost performance further. 
Table~\ref{tbl:baseline_comparison} compares the AccelTran framework with popular transformer accelerators.

\textcolor{black}{We take motivation from SPRING and reuse some hardware modules with minor changes, like the MAC lane (we
add the GeLU activation), the pre-sparsity module, and the post-sparsity module. However, we design many new modules,
namely, specialized RTL modules for the softmax and layer-norm operations, a module to carry out the DynaTran operations
in a single clock cycle, and a novel control block that maps the transformer computational graph to
hardware-implementable tiled operations. The control block is also responsible for choosing among various dataflows,
originally not supported in SPRING. Unlike SPRING, it implements smart scheduling of operations to enable higher
throughput in transformer evaluations (see Section~\ref{sec:smart_scheduling}). This is especially relevant to
transformers with homogeneous operations throughout the model depth. Finally, AccelTran implements a lightweight dynamic inference algorithm for transformers, which SPRING does not support.}

\textcolor{black}{One could evaluate vision transformers (ViTs)~\cite{vit_2021} in AccelTran. However, this would require specialized hardware modules and data-processing pipelines to support image-to-sequence conversion in order to run ViT inference. AccelTran only supports model inference and specialized modules are required to accelerate the backpropagation process in transformer training. We leave these extensions to future work.}

\subsection{Hardware-software Co-design}

In addition to leveraging sparsity in transformers, as explained in Section~\ref{sec:sparsity_sa}, many more techniques 
have been proposed to obtain efficient transformers for pragmatic hardware implementation. These include low-bit 
quantization, knowledge distillation~\cite{turc2019}, approximation of the self-attention 
operation~\cite{linformer, fastformer}, and weight pruning~\cite{compressing_bert, movement_pruning, block_pruning}. 
Further, researchers have proposed hardware-aware neural-architecture search to guide the exploration of efficient 
transformer architectures with hardware feedback~\cite{hat_mit}. However, these works are only limited to certain 
embedded devices~\cite{hat_mit}, FPGAs~\cite{ftrans, fpga_accelerator_1, fpga_accelerator_2}, or off-the-shelf 
microcontrollers~\cite{one_proxy_device} that are far from being optimized for large and compute-heavy transformer 
models. Leveraging the various design decisions in the AccelTran framework can enable efficient and fast 
co-design of the transformer architecture and hardware accelerator. This could incorporate user-defined 
constraints on model accuracy and target power envelopes in diverse deployments~\cite{flexibert}. 
We leave this to future work.

\section{Conclusion}
\label{sec:conclusion}

In this work, we presented AccelTran, a cycle-accurate accelerator simulator that efficiently runs dynamic inference 
with a given transformer model. We proposed a novel, low-overhead dynamic inference scheme, DynaTran, that increases 
the sparsity of activations at runtime with controllable accuracy loss. DynaTran achieves higher accuracy than the 
state-of-the-art top-$k$ hardware-aware pruning strategy while enabling up to 1.33$\times$ higher sparsity. We 
further implement this method on two accelerator architectures: AccelTran-Edge and AccelTran-Server, specialized for 
mobile and cloud platforms, respectively. AccelTran-Edge achieves 330K$\times$ higher throughput at 93K$\times$ 
lower energy when compared to a Raspberry Pi device. Finally, AccelTran-Server achieves 5.73$\times$ higher throughput 
and 3.69$\times$ lower energy consumption relative to the state-of-the-art transformer co-processor, Energon. 

\section*{Acknowledgments}

The simulations presented in this article were performed on computational resources managed and supported by 
Princeton Research Computing at Princeton University.




\bibliographystyle{IEEEtran}
{\footnotesize
\bibliography{IEEEabrv, biblio}}

\begin{IEEEbiography}[{\includegraphics[width=1in,height=1.5in,clip,keepaspectratio]{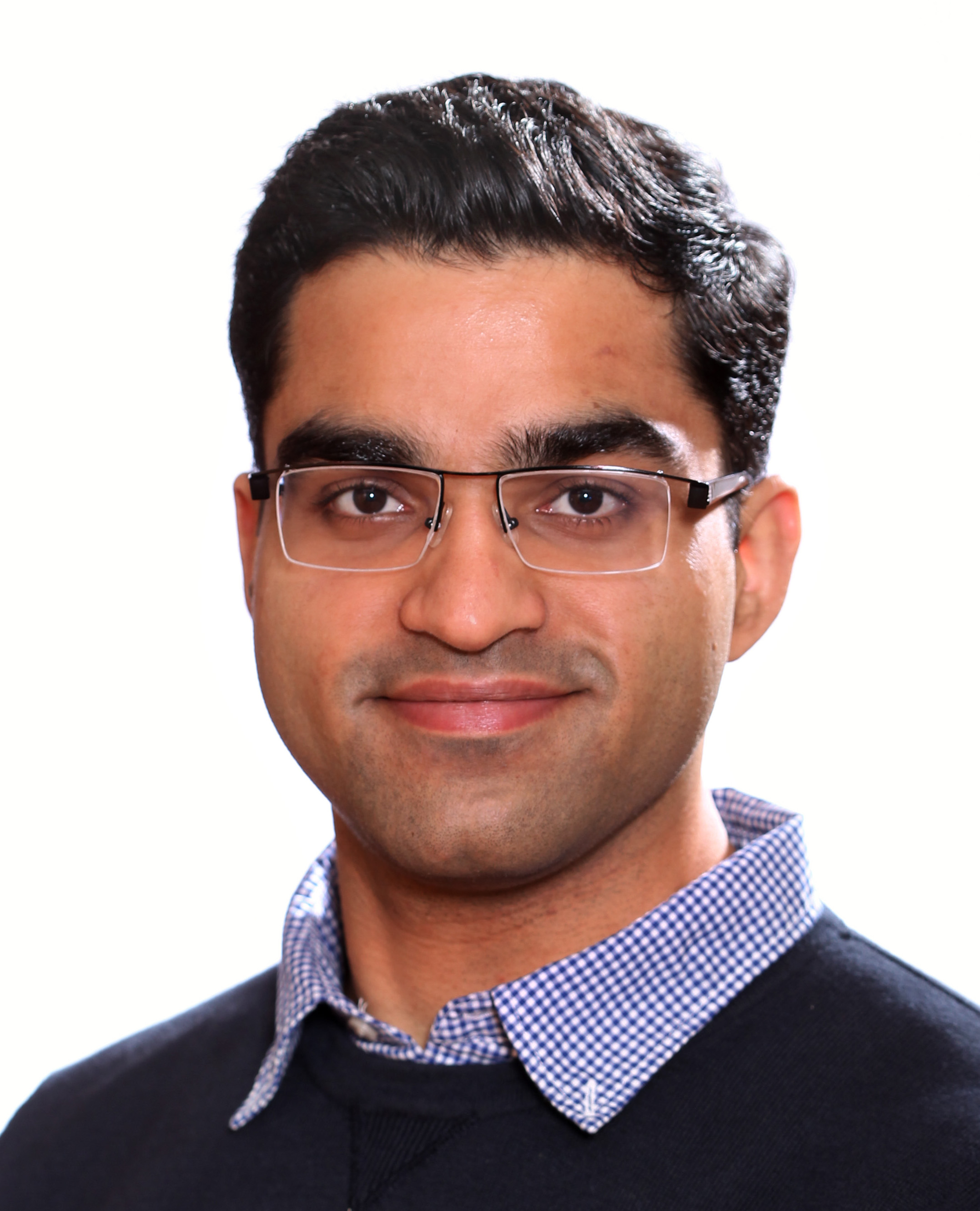}}]{Shikhar Tuli}
received the B. Tech. degree in electrical and electronics engineering from the Indian Institute of Technology (IIT) 
Delhi, India, with a department specialization in very large-scale integration (VLSI) and embedded systems. He is 
currently pursuing a Ph.D. degree at Princeton University in the department of electrical and computer engineering. 
His research interests include deep learning, edge artificial intelligence (AI), hardware-software co-design, 
brain-inspired computing, and smart healthcare.
\end{IEEEbiography}

\begin{IEEEbiography}[{\includegraphics[width=1in,height=1.5in,clip,keepaspectratio]{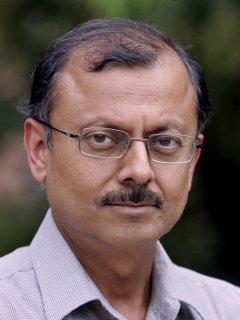}}]{Niraj K. Jha}
(Fellow, IEEE) received the B.Tech. degree in electronics and electrical communication engineering from 
IIT, Kharagpur, India, in 1981, and the Ph.D. degree in electrical engineering from the 
University of Illinois at Urbana–Champaign, Champaign, IL, USA, in 1985. 
He is a professor of electrical and computer engineering, Princeton University. 
He has co-authored five widely used books. He has published more than 470 papers (h-index: 82). 
He has received the Princeton Graduate Mentoring Award. His research has won 15 best paper awards, six award 
nominations, and 25 patents. He was given the Distinguished Alumnus Award by IIT, Kharagpur, in 2014. He has served 
as the Editor-in-Chief of TVLSI and an associate editor of several IEEE Transactions and other journals. He has 
given several keynote speeches in the areas of nanoelectronic design/test, smart healthcare, and cybersecurity.  
He is a Fellow of ACM. His research interests include smart healthcare and machine learning algorithms/architectures. 
\end{IEEEbiography}

\end{document}